\documentclass[preprint]{aastex61}

\usepackage{subfigure}
\usepackage{url}
\usepackage{comment}
\usepackage{color}
\usepackage{enumitem}
\usepackage{amsmath}

\setlist[enumerate]{noitemsep}
\setlist[enumerate,1]{leftmargin=*}
\setlist[itemize]{noitemsep}
\setlist[itemize,1]{leftmargin=*}
\setlist[description]{noitemsep}
\setlist[description,1]{leftmargin=*}
\setenumerate[0]{label=(\arabic*)}

\shorttitle{The RPA: A New Low-$\alpha$, $r$-I, Metal-Poor Halo Star}
\shortauthors{Sakari et al.}

\begin{document}

\newcommand{\target}{J0937$-$0626}
\newcommand{\longtarget}{RAVE J093730.5$-$062655}
\newcommand{\targetsp}{J0937$-$0626 }
\newcommand{\RetIIStar}{DES J0335$-$5403}
\newcommand{\RetIIStarsp}{DES J0335$-$5403 }

\title{The $R$-Process Alliance: Discovery of a low-$\alpha$, $r$-Process-Enhanced Metal-Poor Star in the Galactic Halo} 

\author{Charli M. Sakari}
\affil{Department of Astronomy, University of Washington, Seattle WA
98195-1580, USA}

\author{Ian U. Roederer}
\affil{Department of Astronomy, University of Michigan, 1085 S. University Ave.,
 Ann Arbor, MI 48109, USA}
\affil{Joint Institute for Nuclear Astrophysics Center for the Evolution of the 
Elements (JINA-CEE), USA}

\author{Vinicius M. Placco}
\affil{Department of Physics, University of Notre Dame, Notre Dame, IN 46556,  USA}
\affil{Joint Institute for Nuclear Astrophysics Center for the
  Evolution of the Elements (JINA-CEE), USA}

\author{Timothy C. Beers}
\affil{Department of Physics, University of Notre Dame, Notre Dame, IN 46556,  USA}
\affil{Joint Institute for Nuclear Astrophysics Center for the Evolution of the 
Elements (JINA-CEE), USA}

\author{Rana Ezzeddine}
\affil{Joint Institute for Nuclear Astrophysics Center for the Evolution of the 
Elements (JINA-CEE), USA}
\affil{Department of Physics and Kavli Institute for Astrophysics and
  Space Research, Massachusetts Institute of Technology, Cambridge, MA
  02139, USA}

\author{Anna Frebel}
\affil{Department of Physics and Kavli Institute for Astrophysics and
  Space Research, Massachusetts Institute of Technology, Cambridge, MA
  02139, USA}
\affil{Joint Institute for Nuclear Astrophysics Center for the Evolution of the 
Elements (JINA-CEE), USA}

\author{Terese Hansen}
\affil{Mitchell Institute for Fundamental Physics and Astronomy and
  Department of Physics and Astronomy, Texas A\&M University, College
  Station, TX 77843-4242, USA}


\author{Christopher Sneden}
\affil{Department of Astronomy and McDonald Observatory, The University of Texas, Austin, TX 7
8712, USA}

\author{John J. Cowan}
\affil{Homer L. Dodge Department of Physics and Astronomy, University of Oklahoma, Norman, OK 73019, USA}

\author{George Wallerstein}
\affil{Department of Astronomy, University of Washington, Seattle WA
98195-1580, USA}

\author{Elizabeth M. Farrell}
\affil{Department of Astronomy, University of Washington, Seattle WA
98195-1580, USA}

\author{Kim A. Venn}
\affil{Department of Physics and Astronomy, University of Victoria,
  Victoria, BC, Canada}

\author{Gal Matijevi\v{c}}
\affil{Leibniz Institut f\"{u}r Astrophysik Potsdam (AIP), An der
Sterwarte 16, 14482 Potsdam, Germany}

\author{Rosemary F.G. Wyse}
\affil{Physics and Astronomy Department, Johns Hopkins University, 3400 North Charles Street, Baltimore, MD 21218, USA}

\author{Joss Bland-Hawthorn}
\affil{Sydney Institute for Astronomy, School of Physics A28, University of Sydney, NSW 2006, Australia}
\affil{ARC Centre of Excellence for All Sky Astrophysics (ASTRO-3D), Australia}

\author{Cristina Chiappini}
\affil{Leibniz Institut f\"{u}r Astrophysik Potsdam (AIP), An der
Sterwarte 16, 14482 Potsdam, Germany}

\author{Kenneth C. Freeman}
\affil{Research School of Astronomy \& Astrophysics, The Australian National University, Cotter Road, Canberra, ACT 2611}

\author{Brad K. Gibson}
\affil{E.A. Milne Centre for Astrophysics, University of Hull, Hull,
  HU6 7RX, United Kingdom}

\author{Eva K. Grebel}
\affil{Astronomisches Rechen-Institut, Zentrum f\"ur
Astronomie der Universit\"at Heidelberg, M\"onchhofstr.\ 12--14,
69120 Heidelberg, Germany}

\author{Amina Helmi}
\affil{Kapteyn Astronomical Institute, University of Groningen, P.O. Box 800,
NL-9700 AV Groningen, The Netherlands}

\author{Georges Kordopatis}
\affil{Universit\'e C\^ote d'Azur, Observatoire de la C\^ote d'Azur, CNRS, Laboratoire Lagrange, France}

\author{Andrea Kunder}
\affil{Saint Martin's University, 5000 Abbey Way SE, Lacey, WA 98503 USA}

\author{Julio Navarro}
\affil{Department of Physics and Astronomy, University of Victoria,
  Victoria, BC, Canada}

\author{Warren Reid}
\affil{Department of Physics and Astronomy, Macquarie University, Sydney, NSW 2109, Australia}
\affil{Western Sydney University, Locked bag 1797, Penrith South, NSW 2751, Australia}

\author{George Seabroke}
\affil{Mullard Space Science Laboratory, University College London, Holmbury St Mary, Dorking, RH5 6NT, UK}

\author{Matthias Steinmetz} 
\affil{Leibniz Institut f\"{u}r Astrophysik Potsdam (AIP), An der
Sterwarte 16, 14482 Potsdam, Germany}

\author{Fred Watson}
\affil{Department of Industry, Innovation and Science, 105 Delhi Road, North Ryde, NSW 2113, Australia}

\correspondingauthor{Charli M. Sakari}
\email{sakaricm@u.washington.edu}

\begin{abstract}
A new moderately $r$-process-enhanced metal-poor star, \longtarget,
has been identified in the Milky Way halo as part of an ongoing survey
by the $R$-Process Alliance.  The temperature and surface gravity
indicate that \targetsp is likely a horizontal branch star.  At
$[\rm{Fe/H}] = -1.86$, \targetsp is found to have subsolar [X/Fe]
ratios for nearly every light, $\alpha$, and Fe-peak element.  The low
$[\alpha/\rm{Fe}]$ ratios can be explained by an $\sim0.6$ dex excess
of Fe; \targetsp is therefore similar to the subclass of
``iron-enhanced'' metal-poor stars.  A comparison with Milky Way field
stars at $[\rm{Fe/H}] = -2.5$ suggests that \targetsp was enriched in
material from an event, possibly a Type Ia supernova, that created a
significant amount of Cr, Mn, Fe, and Ni and smaller amounts of Ca,
Sc, Ti, and Zn.  The $r$-process enhancement of \targetsp is likely
due to a separate event, which suggests that its birth environment was
highly enriched in $r$-process elements.  The kinematics of \target,
based on {\it Gaia} DR2 data, indicate a retrograde orbit in the Milky
Way halo; \targetsp was therefore likely accreted from a dwarf galaxy
that had significant $r$-process enrichment.
\end{abstract}


\section{Introduction}\label{sec:Intro}
The advent of large surveys has provided insight into the formation
and evolution of the Milky Way (MW) and its satellites, particularly
the nucleosynthesis of the elements and chemical evolution in galaxies
of different masses.  Many open questions remain, however, including
the astrophysical site for the creation of the heaviest elements in
the Universe.  These elements are created by the rapid ($r$-) neutron
capture process; suggestions that $r$-process nucleosynthesis could
occur during a neutron star merger
\citep{LattimerSchramm1974,Rosswog2014,Lippuner2017} have now been
confirmed through observations of GW~170817
\citep{Abbott2017,Chornock2017,Drout2017,Shappee2017}.  However,
core-collapse supernovae from strongly magnetic stars (the so-called
``jet-supernovae'') may also be a viable site of the $r$-process
(e.g., \citealt{Winteler2012}, \citealt{Cescutti2015},
\citealt{Cote2018}).  One of the most useful sites for probing the
environments, yields, and occurrence rates for $r$-process
nucleosynthesis are the $r$-process-enhanced metal-poor stars, which
retain a relatively pure $r$-process signature and whose spectra are
not overly contaminated from metal lines.

A new collaboration, the $R$-Process Alliance (RPA), has begun a
campaign to identify more of these $r$-process-enhanced metal-poor
stars (with $[\rm{Ba/Eu}]<0$), with the ultimate goal of constraining
the site(s) of the $r$-process across cosmic time.  Initial results
from the Northern and Southern hemisphere surveys
(\citealt{Sakari2018b,Hansen2018}, plus additional papers from
\citealt{Placco2017}, \citealt{Sakari2018a}, \citealt{Cain2018},
\citealt{Gull2018}, \citealt{Holmbeck2018}, and
\citealt{Roederer2018b}) have identified many more of these stars,
including 18 new highly-enhanced $r$-II stars (with
$[\rm{Eu/Fe}]>+1.0$) and 101 new moderately-enhanced $r$-I stars (with
$+0.3\le[\rm{Eu/Fe}]\le+1.0$), according to the classifications from
\citet{BeersChristlieb2005}. These new discoveries enable the
$r$-process-enhanced metal-poor stars to be studied as stellar
populations, so that their chemical and kinematic properties can be
assessed as a whole.

Though they serve as useful laboratories for studying the $r$-process,
it is still not known how or where $r$-process-enhanced stars form,
including how they have retained such a strong $r$-process signal
without being significantly diluted by the nucleosynthetic products of
other stars (e.g., core collapse supernovae).  One theory is
that the $r$-process-enhanced stars form in the lower mass ultra-faint
dwarfs which are later accreted into the MW halo.  This framework is
supported by both observations and simulations: $r$-process-enhanced
stars have been found in ultra faint dwarfs, notably Reticulum~II
\citep{Ji2016,Roederer2016}, while simulations suggest that low mass
dwarf galaxies are capable of retaining the ejecta from an $r$-process
nucleosynthetic event (e.g., \citealt{BlandHawthorn2015},
\citealt{Beniamini2018}).  In addition, the $r$-process-enhanced stars
are also predominantly old (e.g.,
\citealt{Placco2017,Holmbeck2018,Sakari2018b,Valentini2018}), and
simulations indicate that many of the oldest stars in the MW halo may
have been accreted (e.g.,
\citealt{SteinmetzMuller1994,Brook2007,Brook2012,ElBadry2018}).

Another convincing piece of evidence for an extragalactic origin for
the $r$-process-enhanced stars comes from kinematics.  Several $r$-II
stars have orbits consistent with accretion from a satellite
\citep{Roederer2018}, while many of the highly enhanced $r$-II and
$r$-I stars have retrograde orbits in the MW halo which indicate an
extragalactic origin (e.g., \citealt{Sakari2018b}).  An increased
number of $r$-I and $r$-II stars, 
combined with increasingly better data from {\it Gaia}
\citep{GaiaREF}, will enable detailed orbits and more subgroups to be
identified, as was done in \citet{Koppelman2018} and
\citet{Roederer2018}.

Some dwarf galaxy stars can also be identified chemically, as a result
of differing chemical evolution in massive and low-mass galaxies
(see, e.g., \citealt{Tolstoy2009}).  This is generally only possible
for intermediate-metallicity stars that have formed after several
previous generations of stars (i.e., after enough time has passed to
allow chemical evolution to proceed differently in the low-mass
environment).  This also requires that the accreted dwarf galaxy
experienced extended epochs of star formation, rather than a single
burst (see \citealt{Webster2015} for evidence that this is possible,
even in the lowest mass ultra faint dwarfs).  The majority of
metal-poor stars are unlikely to show the chemical signatures of more
metal-rich dwarf galaxy stars.  A few exceptions have been identified
in the MW halo, notably the class of ``Fe-enhanced'' metal-poor stars
which show low [X/Fe] ratios at $[\rm{Fe/H}]<-1$ (e.g.,
\citealt{Yong2013}); generally, however, these stars are fairly
rare. Until now, none of these stars in the MW halo have been
$r$-process-enhanced.

This paper reports the discovery of an $r$-process-enhanced,
metal-poor star that exhibits the typical chemical signatures of dwarf
galaxy stars (notably low [$\alpha$/Fe] ratios).  Section
\ref{sec:Observations} describes the observations, data reduction, and
atmospheric parameters of this star, while Section \ref{sec:Abunds}
presents the abundances.  The implications of these abundances, the
kinematics, and comparisons with other MW halo stars and dwarf galaxy
stars are discussed in Section \ref{sec:Discussion}.

\section{Observations, Data Reduction, and Atmospheric Parameters}\label{sec:Observations}
\targetsp was identified as a metal-poor star in Data Release 4 of the
RAdial Velocity Experiment (RAVE; \citealt{RAVEref,RAVEDR4ref}) and
the subsequent re-analysis by \citet{Matijevic2017}.  It was
then targeted for a medium-resolution, optical analysis by
\citet{Placco2018}.  \targetsp was then observed at high spectral
resolution in 2016 and 2017 using the Astrophysical Research
Consortium (ARC) 3.5-m telescope at Apache Point Observatory, as part
of the Northern Hemisphere survey of the RPA \citep{Sakari2018b}. The
ARC Echelle Spectrograph was used in its default mode, leading to a
spectral resolution of $R\sim31,500$ and coverage of nearly the full
optical range, from $3800$ to $10400$ \AA.  The exposure times were
selected to ensure high S/N ratios in the red and the blue, as shown
in Table \ref{table:Targets}.  The data were reduced in the Image
Reduction and Analysis Facility program (IRAF)\footnote{IRAF is
  distributed by the National Optical Astronomy Observatory, which is
  operated by the Association of Universities for Research in
  Astronomy, Inc., under cooperative agreement with the National
  Science Foundation.} using standard techniques, as described in
\citet{Sakari2018b}.  The heliocentric radial velocity was found by
cross-correlating the spectrum with a high-resolution, high-S/N
spectrum of Arcturus \citep{Hinkle2003}.  The radial velocity is in
excellent agreement with the value from RAVE DR5 (see Table
\ref{table:Targets}).

\begin{figure*}[h!]
\begin{center}
\hspace*{-0.4in}
\includegraphics[scale=0.5,clip,trim=0.8in 0 1.7in 0.5in]{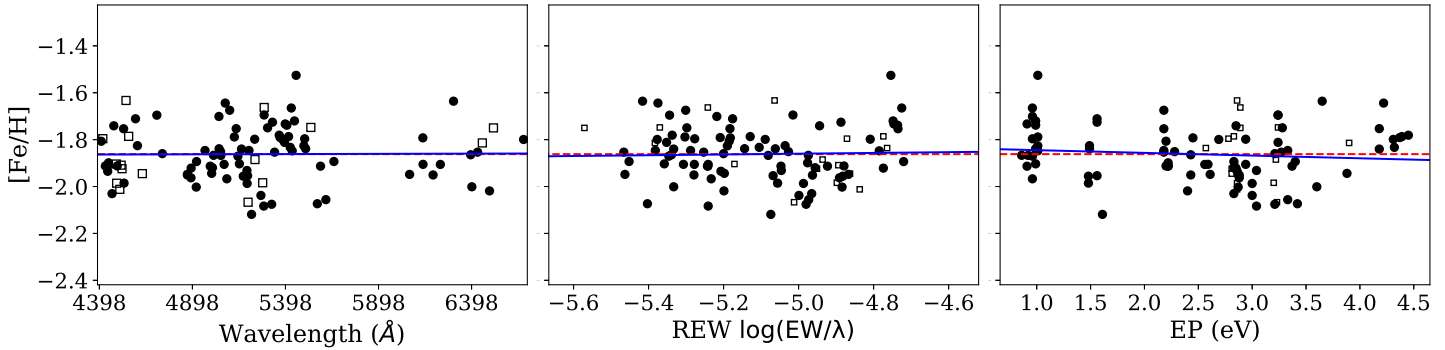}
\caption{Trends in [Fe/H] for \ion{Fe}{1} and \ion{Fe}{2} lines
  (filled circles and open squares, respectively) in \target.  The
  dashed red line shows the average \ion{Fe}{1} abundance, while the
  solid blue lines show the least-square fits to the \ion{Fe}{1} lines.
\label{fig:Trends}}
\end{center}
\end{figure*}

Equivalent widths (EWs) of \ion{Fe}{1} and \ion{Fe}{2} lines from
\citet{Fulbright2006}, \citet{Venn2012}, and \citet{McWilliam2013}
were found using the automated program {\tt DAOSPEC} \citep{DAOSPECref}. 
Fe abundances were then determined using the 2017 version of {\tt
  MOOG} \citep{Sneden}, with an appropriate treatment of scattering
\citep{Sobeck2011}.\footnote{\url{https://github.com/alexji/moog17scat}}
The $<$3D$>$, non-Local Thermodynamic Equilibrium (NLTE) corrections
from \citet{Amarsi2016} were applied to each \ion{Fe}{1} line, as
discussed in \citet{Sakari2018b}.  The temperature and microturbulent
velocity of \targetsp were determined by removing trends in the NLTE
\ion{Fe}{1} abundances with wavelength, reduced EW, and
excitation potential (see Figure \ref{fig:Trends}); the surface
gravity was found by forcing agreement between NLTE \ion{Fe}{1} and
\ion{Fe}{2} abundances.   The final adopted parameters are listed in
Table \ref{table:Targets}, along with the LTE parameters and the
parameters derived with the 1D NLTE corrections of \citet[also see
  \citealt{Ezzeddine2016} and \citealt{Sakari2018b} for more
  details]{Ezzeddine2017}.  The Ezzeddine et al. corrections lead to
similar parameters as the Amarsi et al. $<$3D$>$ NLTE
corrections.  The largest effect of the NLTE corrections to the
\ion{Fe}{1} lines is to raise the surface gravity and the [Fe/H] over
the LTE values.

\citet{Schuster2004} and \citet{Beers2007} obtained photometry of
\target, finding colors that are consistent with the
spectroscopic parameters derived here.  \citet{Schuster2004}
classified \targetsp as a
``red-horizontal-branch-asymptotic-giant-branch transition'' star,
while \citet{Beers2007} found that it was displaced from the
metal-poor main sequence, potentially as a result of its lower surface
gravity.  Indeed, the spectroscopic parameters for \targetsp place it
in the expected region for old, moderately metal-poor HB stars.  The
spectroscopic temperature is also in agreement with the photometric
analysis by \citet{Munari2014} and the spectroscopic RAVE DR5 value
\citep{RAVEDR5ref}, while the temperature, surface gravity, and
metallicity are in agreement with the medium-resolution analysis of
\citet{Placco2018}.  {\it Gaia} has provided a parallax for \targetsp
in Data Release 2 \citep{GaiaREF,GaiaDR2REF}, which gives a
distance; \citet{BailerJones2018} also provide a
statistically-determined distance (see Table \ref{table:Targets}).
These distances, combined with the $E(B-V)$ from the
\citet{SchlaflyFinkbeiner2011} reddening maps, indicate that
\targetsp likely has an absolute magnitude of
$M_{\rm{V}}~=~-0.120\pm0.18$ (with the inverse parallax distance) or
$M_{\rm{V}}~=~0.067\pm0.18$ (with the \citealt{BailerJones2018}
distance).  These magnitudes are both consistent with \targetsp being
a red horizontal branch star.

Very few red horizontal branch stars have been observed by the RPA;
the targets are mainly red giant branch stars.  However, though
\targetsp may be a horizontal branch star, its abundances should
reflect the composition of typical MW stars (with the exception of C;
see the discussion in Section \ref{subsec:Scenarios} and Figure 2 in
\citealt{Roederer2018b}).  Also note that its atmospheric parameters
place \targetsp within the extent of the NLTE grid from \citet[see
  their Table 2]{Amarsi2016}.  To further confirm that the NLTE
corrections are appropriate for a red horizontal branch star, the more
metal-rich $r$-II star HD~222925 from \citet{Roederer2018b} was
re-analyzed.  Its parameters with the $<$3D$>$ NLTE Amarsi et
al. corrections ($T_{\rm{eff}}~=~5625$~K, $\log~g~=~2.3$,
$\xi~=~1.75$~km~s$^{-1}$, and [\ion{Fe}{1}/H]~$=~-1.44$) agree with
the photometric parameters from Roederer et
al. ($T_{\rm{eff}}~=~5636$~K, $\log~g~=~2.54$) and are higher than the
spectroscopic parameters ($\xi~=~2.20$~km~s$^{-1}$, and
[\ion{Fe}{1}/H]~$=~-~1.58$; though note than Roederer et al. find
[\ion{Fe}{2}/H]~$=~-1.47$).  The offsets in the metallicity and
microturbulent velocity are consistent with the general trends found
in LTE vs. NLTE comparison (e.g., \citealt{Amarsi2016}).  However, in
HD~222925 these atmospheric parameter offsets only lead to small
differences in the [X/Fe] ratios ($\la 0.1$ dex); this indicates that
the $<$3D$>$ NLTE \ion{Fe}{1} corrections produce reasonable results
for red horizontal branch stars.

Carbon abundances were found by synthesizing the CH {\it G}-band
region at 4312 \AA.  \targetsp is found to have a subsolar
$[\rm{C/Fe}]~=~-0.55\pm0.40$, a reasonably low value given its
advanced evolutionary state.  Taking the evolutionary corrections of
\citet{Placco2014c} into account, the ``natal'' carbon abundance was
likely higher, at $[\rm{C/Fe}] \sim 0.1$.  \targetsp is not (and never
was) a CEMP star.

\begin{deluxetable}{@{}lcl}
\tabletypesize{\scriptsize}
\tablecolumns{3}
\tablewidth{6in}
\tablecaption{Target Information\label{table:Targets}}
\hspace*{-2in}
\tablehead{
Parameter & Value & Notes}
\startdata
ID                      & \longtarget & Other IDs: TYC 4900-1967-1,
BS 17576-0027, 2MASS J09373053-0626551 \\
RA (J2000)              & 09:37:30.54  & \\
Dec (J2000)             & -06:26:55.0  & \\
$V$                     & 11.81        & \\
$K$                     & 10.13        & \\
$E(B-V)$                & 0.0266       & Average value from \citet{SchlaflyFinkbeiner2011} maps \\
$d$ (kpc)               & $2.342^{+0.230}_{-0.192}$  & Inverse parallax distance \\
                        & $2.149^{+0.190}_{-0.163}$  & \citet{BailerJones2018} \\
$M_{\rm{V}}$              & $-0.120\pm0.18$ & Calculated from the
inverse parallax distance\\
                        & $0.067\pm0.18$ & Calculated from the
\citet{BailerJones2018} distance \\
Observation Dates       & 28 Jan, 11 Feb 2016, 2 Mar 2017 & Seeing =
0.9, 1.09, 1.12\arcsec \\
Exposure Time (s)       & 3240  & \\
S/N, 4400 \AA           & 100  & Per pixel; there are 2.5 pixels per
  resolution element \\
S/N, 6500 \AA           & 170  & Per pixel; there are 2.5 pixels per
  resolution element \\
 & & \\
$v_{\rm{helio}}$ (km s$^{-1}$)    & $268.8\pm1.0$  & This work\\
                              & $268.436\pm1.297$ & RAVE DR5 \\
 & & \\
$T_{\rm{eff}}$ (K)         & $\boldsymbol{5875\pm55}^{a}$ & Spectroscopic, with $<$3D$>$ NLTE correction; this work \\
                         & $5875\pm55$ & Spectroscopic, LTE; this work \\
                         & $5850\pm50$ & Spectroscopic, with 1D NLTE \citet{Ezzeddine2017} corrections; this work \\
                         & $6091$      & Spectroscopic, \citet{Placco2018} \\
                         & $5667.31\pm214$ & Spectroscopic, RAVE DR5 \\
                         & 5606        & Photometric, \citet{RamirezMelendez2005} calibration \\
                         & 5752        & Photometric, \citet{Casagrande2010} calibration \\
$\log g$                 & $\boldsymbol{2.61\pm0.16}^{a}$  & Spectroscopic, this work \\
                         & $2.31\pm0.16$ & Spectroscopic, LTE; this work  \\
                         & $2.70\pm0.20$ & Spectroscopic, with 1D NLTE \citet{Ezzeddine2017} corrections; this work \\
                         & $2.52$        & Spectroscopic, \citet{Placco2018} \\
                         & $2.81\pm0.48$ & Spectroscopic, RAVE DR5 \\
$\xi$ (km/s)            & $\boldsymbol{2.09\pm0.25}^{a}$  & This work \\
                        & $2.14\pm0.25$ & Spectroscopic, LTE; this work  \\
                        & $2.20\pm0.20$ & Spectroscopic, with 1D NLTE \citet{Ezzeddine2017} corrections; this work \\
$[$M/H$]$               & -2.04          & RAVE DR5 \\
$[$Fe/H$]$              & $\boldsymbol{-1.86\pm0.02}^{a}$ & This work \\
                        & $-2.03\pm0.02$ & Spectroscopic, LTE; this work  \\
                        & $-1.89\pm0.15$ & Spectroscopic, with 1D NLTE \citet{Ezzeddine2017} corrections; this work \\
                        & $-1.70$        & \citet{Placco2018} \\
$[$C/Fe$]$              & $\boldsymbol{-0.55\pm0.40}^{a}$  & Measured value, this work \\
                        & $\sim+0.1$       & ``Natal'' value, calculated
with the evolutionary corrections of \citet{Placco2014c}\tablenotemark{b} \\
                        & $+0.38$         & Measured value, \citet{Placco2018} \\
\enddata
\tablenotetext{a}{The bold values show the final spectroscopic
  values adopted for the abundance analysis.}
\tablenotetext{b}{This correction assumes that the star is a
  horizontal branch star, and therefore has the same level of C
  depletion as a tip of the red giant branch star.}
\end{deluxetable}

\section{Detailed Abundances}\label{sec:Abunds}
Abundances of Fe, Ca, Sc, Ti, Cr, Mn, Co, and Ni were determined
from EWs; all other elements were determined from spectrum syntheses.
The lines for the EW analyses are from the line lists of
\citet{Fulbright2006,Fulbright2007} and \citet{McWilliam2013}.
Corrections for hyperfine structure and (if necessary) isotopic
splitting were included for Sc, V, Mn, and Co, using the data from 
the Kurucz
database\footnote{\url{http://kurucz.harvard.edu/linelists.html}} and
\citet{McWilliam2013}.  The spectrum synthesis line lists were
generated with the {\tt linemake}
code.\footnote{\vspace{0.5in}\url{https://github.com/vmplacco/linemake}}
Hyperfine structure, isotopic splitting, and molecular lines from CH,
C$_{2}$, and CN were included in the synthetic spectrum line lists.
All [X/H] ratios were calculated line-by-line, where the solar
abundance has been determined from the Kurucz solar
spectrum\footnote{\url{http://kurucz.harvard.edu/sun.html}} if the
lines are sufficiently weak and unblended, using the same atomic data;
otherwise, the \citet{Asplund2009} solar values are adopted.
Note that unlike \citet{Sakari2018b}, a differential analysis has not
been utilized, because there is not a suitable standard star in this
metallicity range.

Table \ref{table:LineAbunds} shows the line-by-line EWs or, for lines
whose abundances were derived from spectrum syntheses, abundances.
Table \ref{table:Abunds} shows the final mean abundances.  For the
EW-based abundances, the random errors represent the line-to-line
dispersion, with a minimum error in a single line of $0.05-0.1$ dex,
depending on the strength of the line and S/N; for abundances that
were determined via spectrum syntheses, the random errors are based on
the quality of the syntheses. Table~\ref{table:Abunds} also shows the
total error, which is a quadrature sum of the random error and the
systematic error due to uncertainties in the atmospheric parameters.
The systematic errors were determined from the variances and
covariances of the atmospheric parameters, according to the techniques
outlined in \citet{McWilliam2013} and \citet{Sakari2018b}.
Table \ref{table:Abunds} also provides the abundance offsets that
occur if LTE parameters are adopted.  The offsets are all $\la 0.2$
dex; all [X/Fe] ratios relative to \ion{Fe}{2} are negligible.  These
offsets reflect the abundance sensitivities to differences in $\log
g$, microturbulent velocity, and [\ion{Fe}{1}/H].  NLTE corrections
were not applied to elements other than \ion{Fe}{1}; significant NLTE
sensitivities are generally noted below.

\begin{deluxetable}{@{}ccccccc}
\tabletypesize{\scriptsize}
\tablecolumns{7}
\tablewidth{0pt}
\tablecaption{Line Equivalent Widths or Abundances\tablenotemark{a}\label{table:LineAbunds}}
\hspace*{-4in}
\tablehead{
Element & Wavelength & EP & $\log gf$ & EW & $\log \epsilon$ & Flag$^{b}$ \\
 & (\AA) &  (eV) &  & (m\AA) & &   
}
\startdata
\ion{Na}{1} & 5895.92 & 0.00 & -0.180 &      & 4.13 & SYN \\
\ion{Mg}{1} & 4167.28 & 4.34 & -0.745 &      & 5.56 & SYN \\
\ion{Mg}{1} & 4703.00 & 4.34 & -0.670 &      & 5.46 & SYN \\
\ion{Mg}{1} & 5528.41 & 4.34 & -0.480 &      & 5.44 & SYN \\
\ion{Al}{1} & 3944.00 & 0.00 & -0.640 &      & 3.17 & SYN \\
\ion{Al}{1} & 3961.52 & 0.01 & -0.340 &      & 3.24 & SYN \\
\ion{Si}{1} & 3905.52 & 1.91 & -1.090 &      & 5.20 & SYN \\
\ion{Ca}{1} & 4283.01 & 1.89 & -0.220 & 35.7 &      & EW \\
\ion{Ca}{1} & 4289.37 & 1.88 & -0.300 & 29.7 &      & EW \\
\ion{Ca}{1} & 4302.54 & 1.90 &  0.275 & 63.1 &      & EW \\
\enddata
\tablenotetext{a}{Only a portion of this table is shown here to
  demonstrate its form and content. A machine-readable version of the
  full table is available.}
\tablenotetext{b}{A flag of ``SYN'' indicates that the abundance was
  determined via spectrum synthesis; in this case, a $\log \epsilon$
  abundance is given.  ``EW'' indicates that an equivalent width
  analysis was performed, and the measured EW is given instead of an
  abundance.}
\tablenotetext{c}{This line has HFS and/or isotopic splitting.}
\end{deluxetable}

\begin{deluxetable*}{@{}lcDccDcD}
\tabletypesize{\scriptsize}
\tablecolumns{8}
\tablewidth{0pt}
\tablecaption{Mean Abundances and Uncertainties\label{table:Abunds}}
\hspace*{-4in}
\tablehead{
Element          & $N$ & \multicolumn{2}{r}{$\log \epsilon$} &
$\sigma_{\rm{random}}$ & $\sigma_{\rm{Tot}}$\tablenotemark{b} &
\multicolumn{2}{c}{$\;\;\;\;$[X/Fe]\tablenotemark{a}} &
$\sigma_{\rm{Tot}}$\tablenotemark{b} & \multicolumn{2}{r}{\phantom{$-$}$\Delta_{\rm{LTE}}$\tablenotemark{c}}
}
\decimals
\startdata
\ion{Fe}{1} & 84 &  5.64 & 0.01 & 0.05 & -1.86 & 0.05  & -0.17 \\
\ion{Fe}{2} & 20 &  5.64 & 0.03 & 0.11 & -1.86 & 0.11  & -0.11 \\
\ion{Na}{1} &  1 &  4.13 & 0.10 & 0.15 & -0.25 & 0.13  & 0.14 \\
\ion{Mg}{1} &  3 &  5.49 & 0.03 & 0.05 & -0.25 & 0.03  & 0.15 \\
\ion{Al}{1} &  2 &  3.19 & 0.06 & 0.09 & -1.40 & 0.07  & 0.10 \\
\ion{Si}{1} &  1 &  5.30 & 0.20 & 0.23 & -0.35 & 0.21  & 0.14 \\
\ion{Ca}{1} & 17 &  4.30 & 0.02 & 0.04 & -0.18 & 0.02  & 0.17 \\
\ion{Sc}{2} &  7 &  0.86 & 0.02 & 0.09 & -0.43 & 0.05  & 0.01 \\
\ion{Ti}{1} &  6 &  3.13 & 0.02 & 0.05 &  0.04 & 0.03  & 0.16 \\
\ion{Ti}{2} & 25 &  3.05 & 0.02 & 0.09 & -0.04 & 0.05  & 0.01 \\
\ion{Cr}{1} &  7 &  3.83 & 0.04 & 0.08 &  0.05 & 0.05  & 0.16 \\
\ion{Cr}{2} &  4 &  3.90 & 0.04 & 0.11 &  0.12 & 0.05  & 0.01 \\
\ion{Mn}{1} &  1 &  3.12 & 0.10 & 0.11 & -0.45 & 0.10  & 0.17 \\
\ion{Co}{1} &  1 &  2.65 & 0.10 & 0.11 & -0.48 & 0.11  & 0.17 \\
\ion{Ni}{1} &  2 &  4.18 & 0.06 & 0.07 & -0.16 & 0.06  & 0.14 \\
\ion{Zn}{1} &  2 &  2.40 & 0.06 & 0.07 & -0.30 & 0.06  & 0.15 \\
\ion{Sr}{2} &  1 &  0.96 & 0.20 & 0.32 & -0.05 & 0.24  & -0.01 \\
\ion{Y}{2}  &  2 &  0.05 & 0.08 & 0.10 & -0.30 & 0.10  & 0.01 \\
\ion{Zr}{2} &  2 &  0.62 & 0.07 & 0.10 & -0.10 & 0.10  & 0.01 \\
\ion{Ba}{2} &  3 &  0.40 & 0.05 & 0.16 &  0.08 & 0.08  & 0.02 \\
\ion{La}{2} &  3 & -0.35 & 0.03 & 0.08 &  0.41 & 0.08  & 0.01 \\
\ion{Ce}{2} &  1 &  0.07 & 0.05 & 0.09 &  0.35 & 0.08  & 0.02 \\
\ion{Pr}{2} &  1 & -0.44 & 0.10 & 0.12 &  0.70 & 0.12  & 0.02 \\
\ion{Nd}{2} &  2 &  0.14 & 0.09 & 0.11 &  0.58 & 0.11  & 0.02 \\
\ion{Eu}{2} &  3 & -0.49 & 0.06 & 0.12 &  0.85 & 0.08  & 0.01 \\
\ion{Gd}{2} &  1 & -0.04 & 0.05 & 0.11 &  0.70 & 0.08  & 0.01 \\
\ion{Dy}{2} &  1 &  0.19 & 0.10 & 0.12 &  0.95 & 0.12  & 0.01 \\
\ion{Th}{2} &  1 & <-0.79 & & & <1.05     &  & \multicolumn{2}{l}{ } \\
\enddata
\tablenotetext{a}{[Fe/H] is given for \ion{Fe}{1} and \ion{Fe}{2}.}
\tablenotetext{b}{The total error refers to the combination of random
  and systematic errors (where the latter are due to uncertainties in
  the atmospheric parameters), calculated according to Equations A1,
  A4, and A5 in \citet{McWilliam2013}.  Errors in $\log \epsilon$ and
  [X/Fe] are listed separately.}
\tablenotetext{c}{$\Delta_{\rm{LTE}}$ shows the offsets in
[X/Fe] ratios that occur when the LTE atmospheres are used.}
\end{deluxetable*}

\subsection{Light Elements: Na and Al}\label{subsec:Light}
The Na abundance was determined from a synthesis of the 5895 \AA $\;$
\ion{Na}{1} line (the 5889 \AA $\;$ line is too strong), indicating a
slightly subsolar [Na/Fe] ratio.  The \ion{Al}{1} lines at 3944 and
3961 \AA $\;$ yield a significantly subsolar
$[\rm{Al/Fe}]~=~-1.40\pm0.06$.  Table \ref{table:Abunds} shows the LTE
abundances, but both the Na and Al lines likely suffer from NLTE
effects.  The {\tt INSPECT}
database\footnote{\url{http://inspect-stars.com/}} \citep{Lind2011}
indicates that the 5895 \AA $\;$ \ion{Na}{1} line should have a NLTE
correction of -0.41 dex, which would make the Na abundance
significantly subsolar.   NLTE corrections to the Al lines may be as
large as $+0.6$ to $0.8$~dex in this temperature, surface gravity, and
metallicity range \citep{NordlanderLind2017}.

\subsection{$\alpha$ Elements}\label{subsec:Alphas}
\begin{figure}[h!]
\begin{center}
\centering
\hspace*{-0.3in}
\includegraphics[scale=0.6,clip,trim=0.5in 0 0 0]{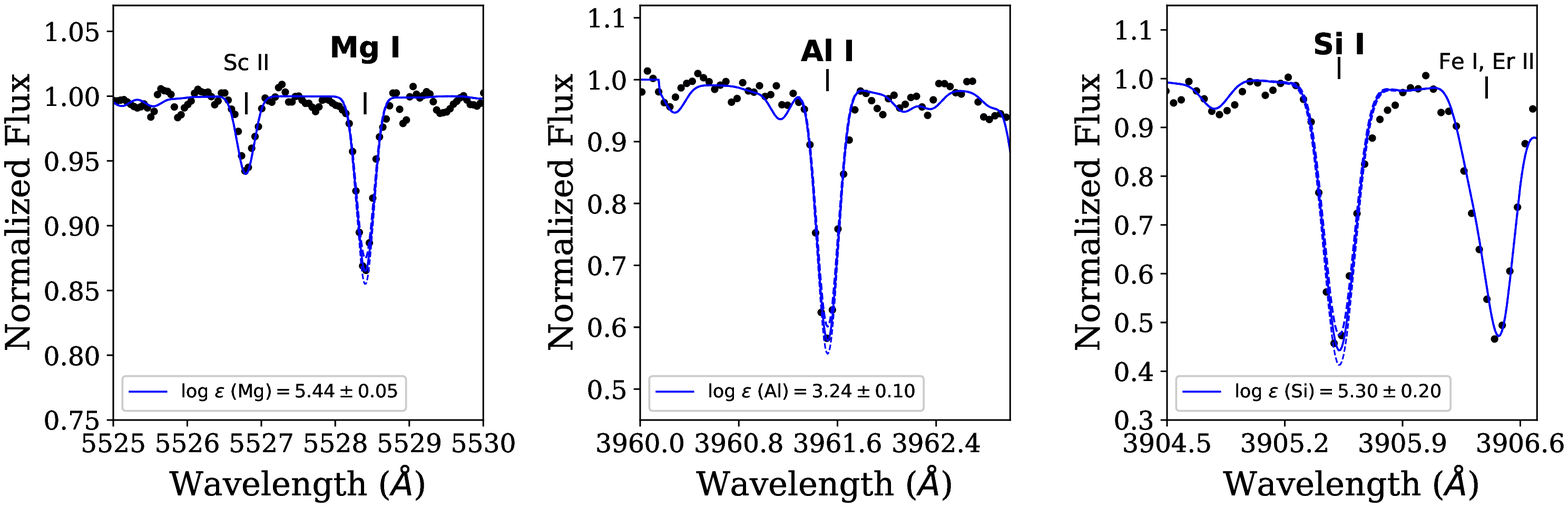}
\caption{Syntheses of Mg, Al, and Si lines.  The solid line shows the
  best-fit, while the dashed lines show the $1\sigma$ uncertainties.
\label{fig:AlphaSynths}}
\end{center}
\end{figure}

The $\alpha$-elements with detectable lines in \targetsp include Mg,
Si, Ca, and Ti; the O lines are too weak.  The three \ion{Mg}{1} lines
at 4167, 4703, and 5528 \AA $\;$ and the \ion{Si}{1} line at 3905 \AA
$\;$ were synthesized (see Figure \ref{fig:AlphaSynths}), while EWs
were measured for 17 \ion{Ca}{1} lines, 6 \ion{Ti}{1} lines, and 25
\ion{Ti}{2} lines.  \citet{Mashonkina2016} show that NLTE corrections
$\sim 0.1$ dex may be required for Ca, with larger and smaller
corrections for \ion{Ti}{1} and \ion{Ti}{2}, respectively (though note
that the highest temperature they consider is 5000 K).  The [X/Fe]
ratios for Mg, Si, and Ca are subsolar, while [Ti/Fe] is roughly
solar; these ratios are subsolar even when LTE parameters are used.
\citet{Placco2018} also found a low $[\alpha/\rm{Fe}]~=~-0.09$ based
on a medium-resolution spectrum.

\subsection{Iron-Peak Elements and Zinc}\label{subsec:FePeak}
EWs were measured for 7 \ion{Sc}{2}, 7 \ion{Cr}{1}, 4
\ion{Cr}{2}, 1 \ion{Mn}{1}, 1 \ion{Co}{1}, and 3 \ion{Ni}{1} lines.
The [\ion{Cr}{1}/Fe] ratio is expected to suffer from small NLTE
effects; \citep{BergemannCescutti2010} find corrections $<0.2$ dex.
The \ion{Mn}{1} and \ion{Co}{1} lines also require NLTE corrections on
the order of $+0.4$ \citep{BergemannGehren2008} and $+0.6$ dex
\citep{Bergemann2010}, respectively, according the MPIA NLTE correction
database\footnote{\url{http://nlte.mpia.de}} (though note that none of
the models extend to \target's surface gravity).  None of these NLTE
corrections were applied.  The \ion{Zn}{1} lines at 4722 and 4810 \AA
$\;$ were synthesized.  The LTE [Sc/Fe], [Mn/Fe], [Co/Fe], and [Ni/Fe]
ratios are all subsolar, [Cr/Fe] is slightly enhanced, and [Zn/Fe] is
slightly subsolar.

\begin{figure}[h!]
\begin{center}
\centering
\hspace*{-0.3in}
\includegraphics[scale=0.6,clip,trim=0.5in 0 0 0]{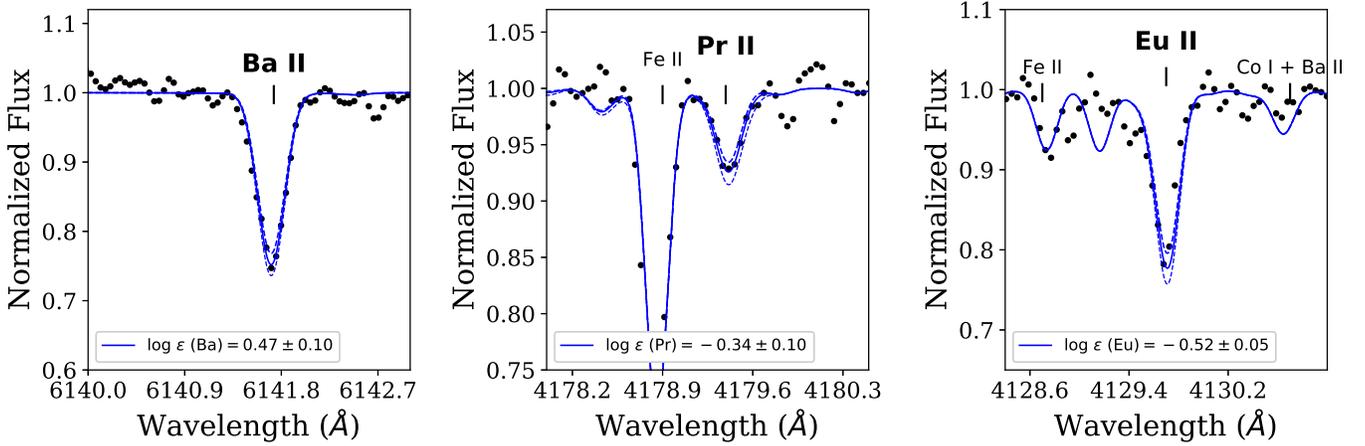}
\caption{Syntheses of Ba, Pr, and Eu lines.
\label{fig:NCSynths}}
\end{center}
\end{figure}

\subsection{Neutron-Capture Elements}\label{subsec:NeutronCaptures}
The Sr abundance in \targetsp was derived from the relatively strong
4215 \AA $\;$ line, Y was derived from the weak 4883 and 4900 \AA $\;$
lines, and Zr was derived from the 4161 and 4208 \AA $\;$
lines. Unlike the lighter elements, Sr and Zr yield
approximately solar [X/Fe] ratios; Y is slightly subsolar.

Barium and europium are the elements used for classification of r-I
and r-II stars.  The 5853, 6141, and 6496 \AA \hspace{0.025in}
\ion{Ba}{2} lines were used (the 4554 \AA $\;$ line is too strong),
while the 3819, 4129, and 4205~\AA \hspace{0.025in} \ion{Eu}{2} lines
were used.  \targetsp has a roughly solar $[\rm{Ba/Fe}] = 0.08\pm0.05$
but an enhanced $[\rm{Eu/Fe}]~=~0.85\pm 0.06$, making it an $r$-I
star. Its low $[\rm{Ba/Eu}]~=~-0.77\pm0.07$ indicates that it has
received minimal contamination from the main $s$-process.  Lines of
\ion{Ce}{2}, \ion{Pr}{2}, \ion{Nd}{2}, \ion{Gd}{2}, and \ion{Dy}{2}
are detectable and also indicate enhancement.  Figure
\ref{fig:NCSynths} shows syntheses of Ba, Pr, and Eu lines.  An upper
limit is derived from the \ion{Th}{2} line at
4019~\AA \hspace{0.025in}.


\section{Discussion}\label{sec:Discussion}
Relative to Fe, \targetsp has subsolar [X/Fe] ratios for nearly all
elements (with the exception of Cr and possibly Ti) other than the
neutron-capture elements. It has approximately solar [X/Fe] ratios for
Sr, Y, Zr, and Ba, yet is enhanced in La, Ce, Pr, Nd, Eu, Gd, and
Dy. Below, \target's abundance patterns are compared to other MW halo
stars and dwarf galaxy stars (Section \ref{subsec:MWComp}), including
the subset of ``Fe-enhanced'' metal-poor stars.  Scenarios to explain
the light, $\alpha$, and Fe-peak elements are explored in Section
\ref{subsec:Scenarios}. The $r$-process enhancement and patterns in
\targetsp are discussed in Section \ref{subsec:Patterns}, while the
kinematics are discussed in Section \ref{subsec:Kinematics}.

\subsection{Comparisons with Milky Way and Dwarf Galaxy Stars}\label{subsec:MWComp}

\begin{figure}[h!]
\begin{center}
\centering
\hspace*{-0.45in}
\subfigure{\includegraphics[scale=0.58,trim=0.02in 0.1in 0.55in 0.4in,clip]{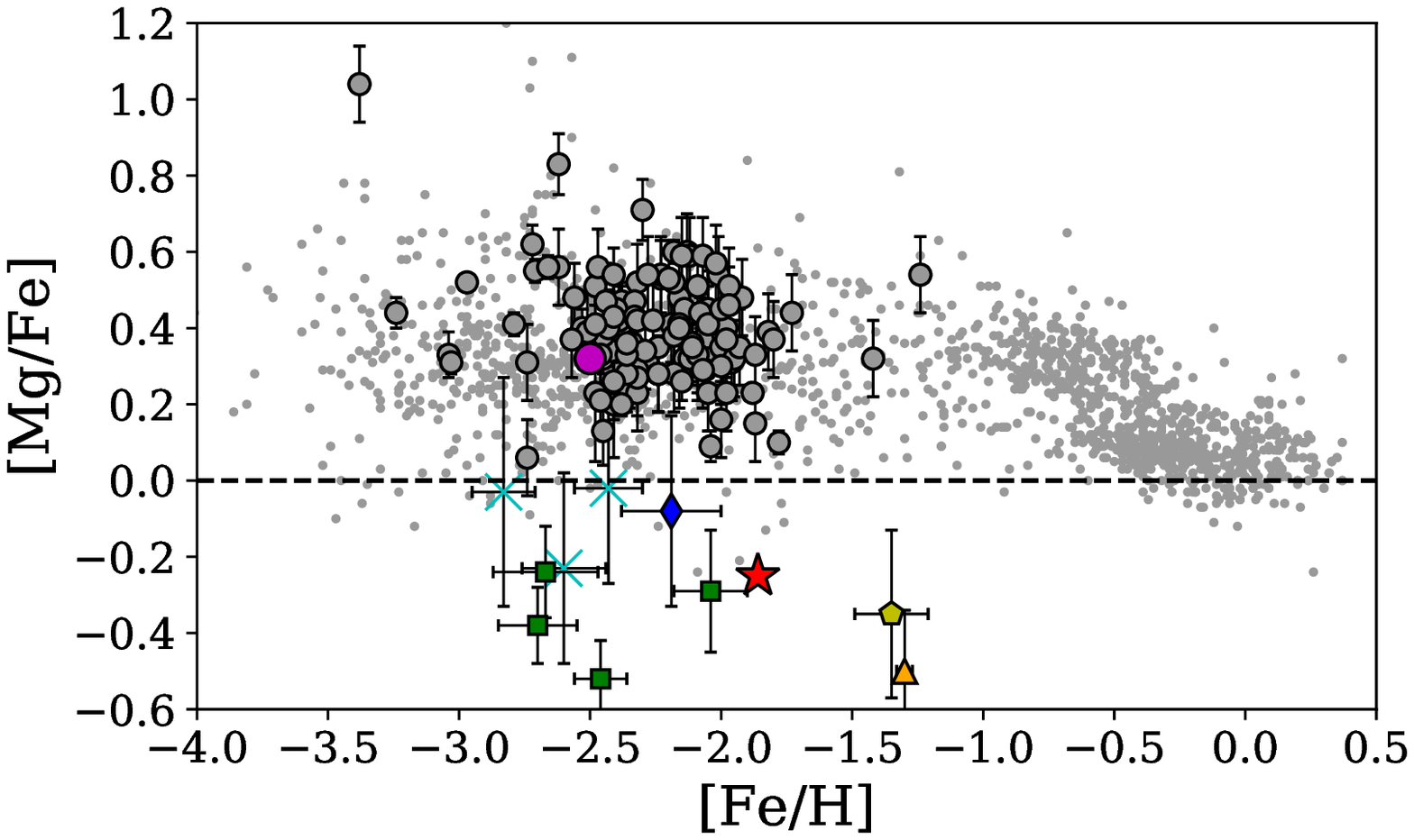}\label{subfig:Mg}}
\subfigure{\includegraphics[scale=0.58,trim=0.02in 0.1in 0.55in 0.4in,clip]{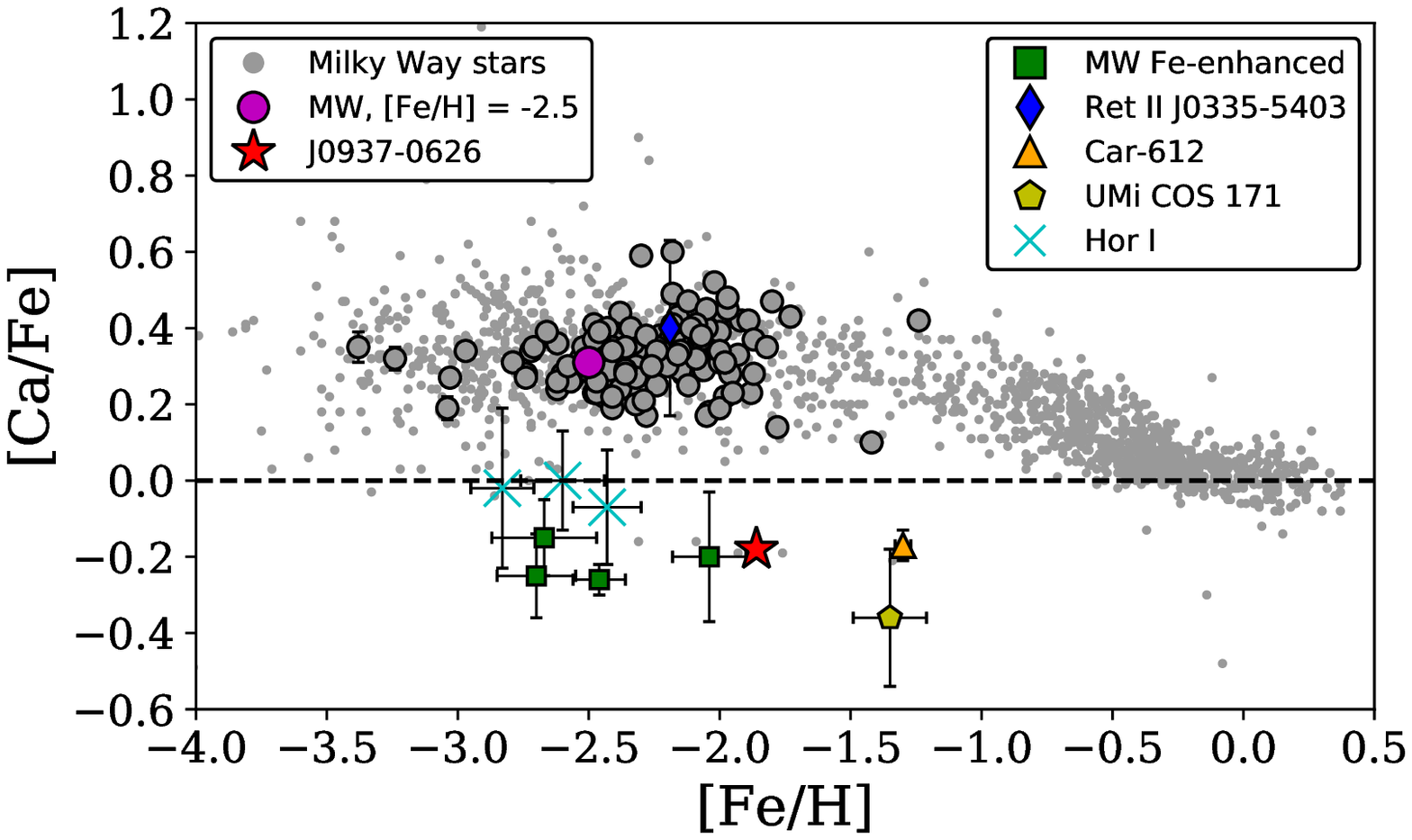}\label{subfig:Ca}}
\caption{[Mg/Fe] and [Ca/Fe] ratios as a function of [Fe/H].
  \target's abundances and uncertainties are shown with a red star.
  The MW stars are shown as grey points \citep{Venn2004,Reddy2006};
  the stars from \citet{Sakari2018b} are the larger, outlined,
  circles.  The MW Fe-enhanced stars from \citet{Yong2013},
  \citet{Cohen2013}, and \citet{Roederer2014b} are shown with green
  squares.  Stars from four dwarf galaxies are shown: Ret~II
  DES~J0335$-$5403, with blue diamonds \citep{Ji2016}; UMi~COS~171,
  with a yellow pentagon \citep{CohenHuang2010}; Car-612, with an
  orange triangle \citep{Venn2012}; and three stars from Hor~I, with
  cyan crosses \citep{Nagasawa2018}.  The magenta circle shows the
  average of the \citet{Sakari2018b} stars with
  $-2.6~\le~[\rm{Fe/H}]~\le~-2.4$.\label{fig:alphaFe}}
\end{center}
\end{figure}

\begin{figure}[h!]
\begin{center}
\centering
\hspace*{-0.45in}
\subfigure{\includegraphics[scale=0.58,trim=0.02in 0.1in 0.55in 0.4in,clip]{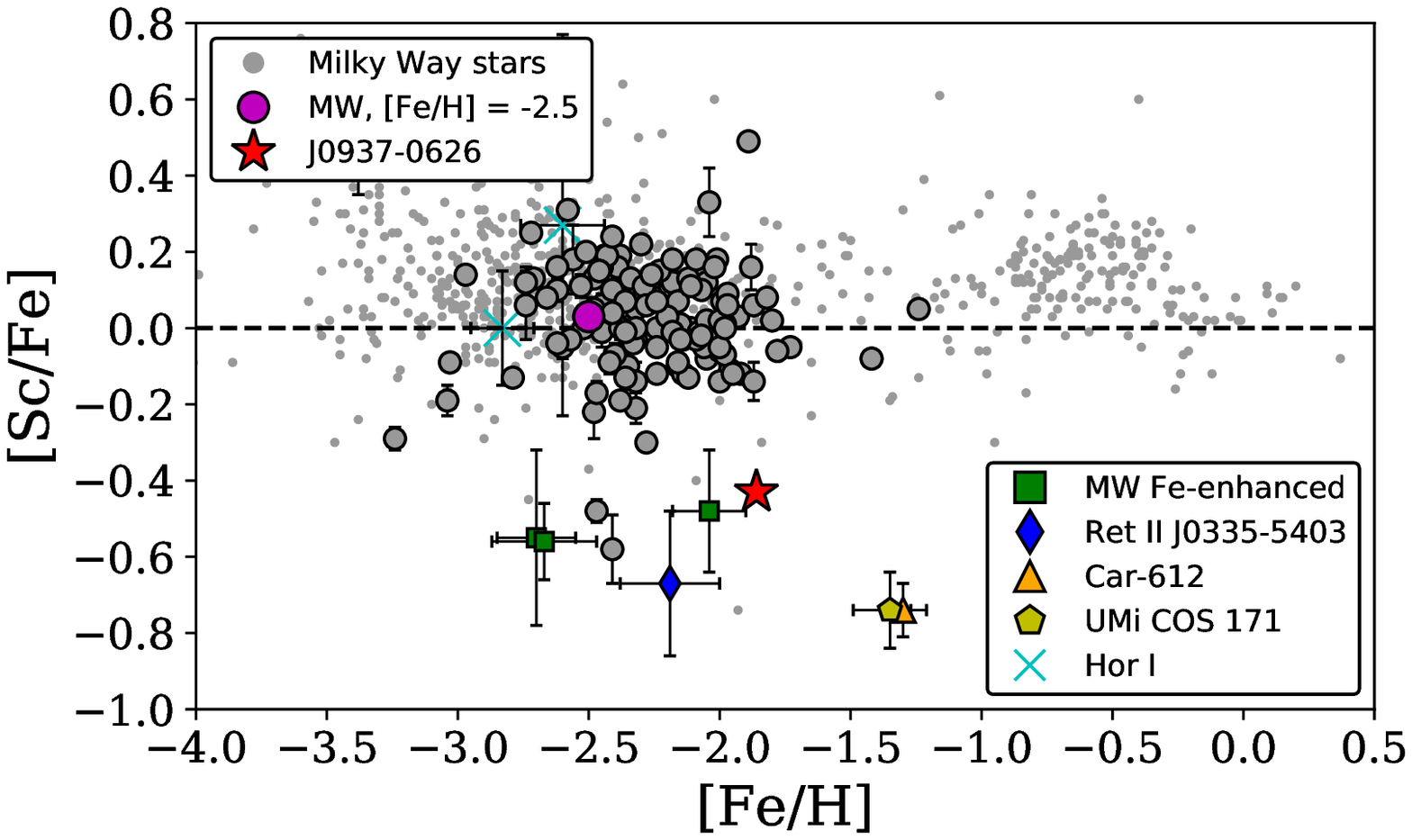}\label{subfig:Sc}}
\subfigure{\includegraphics[scale=0.58,trim=0.02in 0.1in 0.55in 0.4in,clip]{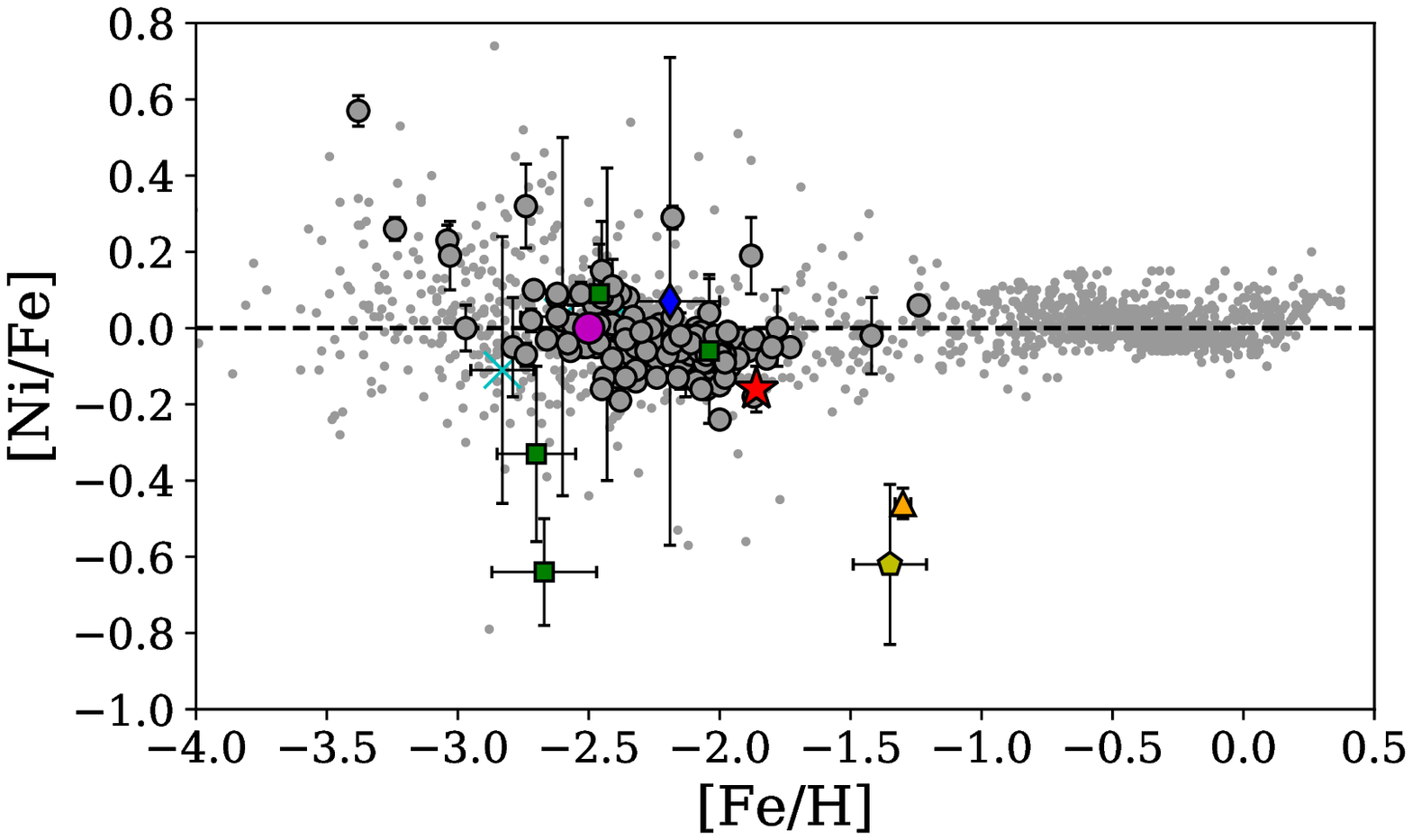}\label{subfig:Mn}}
\caption{[Sc/Fe] and [Ni/Fe] ratios as a function of [Fe/H].  Points
  are as in Figure \ref{fig:alphaFe}.
\label{fig:FePeakFe}}
\end{center}
\end{figure}

Figure \ref{fig:alphaFe} demonstrates that \targetsp has low [Mg/Fe]
and [Ca/Fe] relative to MW field stars, along with [Si/Fe] and
[Ti/Fe]. \targetsp therefore has a deficiency of $\alpha$-elements,
relative to Fe, compared to MW field stars.  This is a phenomenon
usually seen in dwarf galaxy stars (e.g.,
\citealt{Shetrone2003,Tolstoy2009}), where lower $[\alpha/\rm{Fe}]$ at
a given $[\rm{Fe/H}]$ is usually interpreted as 
a sign of enrichment from Type Ia supernovae which produce lots of Fe,
but few $\alpha$-elements (e.g.,
\citealt{Lanfranchi2008}).\footnote{Note that \targetsp is
more iron- and $\alpha$-poor than the low-$\alpha$ disk stars (e.g.,
\citealt{Helmi2018}), as will be discussed in Section
\ref{subsec:Kinematics}.}  However, with the exception of the
neutron-capture elements (which will be discussed in Section
\ref{subsec:Patterns}), most of \target's [X/Fe] ratios are lower than
MW field stars at the same metallicity, including Sc and Ni (Figure
\ref{fig:FePeakFe}; though note that Mn and Cr agree with MW stars).
In this respect, \targetsp resembles the class of MW and dwarf galaxy
stars that have been called ``Fe-enhanced'' metal-poor stars
\citep{Yong2013}, stars which show subsolar [X/Fe] ratios in
nearly every element.\footnote{The low Sc in \targetsp is also
  somewhat reminiscent of the metal-poor bulge stars observed by
  \citet{CaseySchlaufman2015}, though those stars have otherwise
  normal [X/Fe] ratios.}  Also shown in Figures \ref{fig:alphaFe} and
\ref{fig:FePeakFe} are a selection of low-$\alpha$ or ``Fe-enhanced''
metal-poor stars from the MW halo, HE~1207$-$3108 \citep{Yong2013},
HE~0533$-$5340 \citep{Cohen2013}, SDSS~J001820.5-093939.2
\citep{Aoki2014}, and BD$+80^{\circ}$~245
\citep{Carney1997,Ivans2003,Roederer2014b}, along with low-$\alpha$ 
stars from four dwarf galaxies, Ursa Minor (UMi COS~171;
\citealt{CohenHuang2010}), Carina (Car 612; \citealt{Venn2012}),
Horologium~I (Hor~I, three stars; \citealt{Nagasawa2018}), and Ret~II
(DES~J033548$-$540349, hereafter \RetIIStar; \citealt{Ji2016}). These
comparison stars indeed show the characteristic subsolar [X/Fe] ratios
found in \target, albeit with slightly different values.  \targetsp
can therefore be considered to be another one of these ``Fe-enhanced''
metal-poor stars.

\begin{figure}[h!]
\begin{center}
\centering
\includegraphics[scale=0.8]{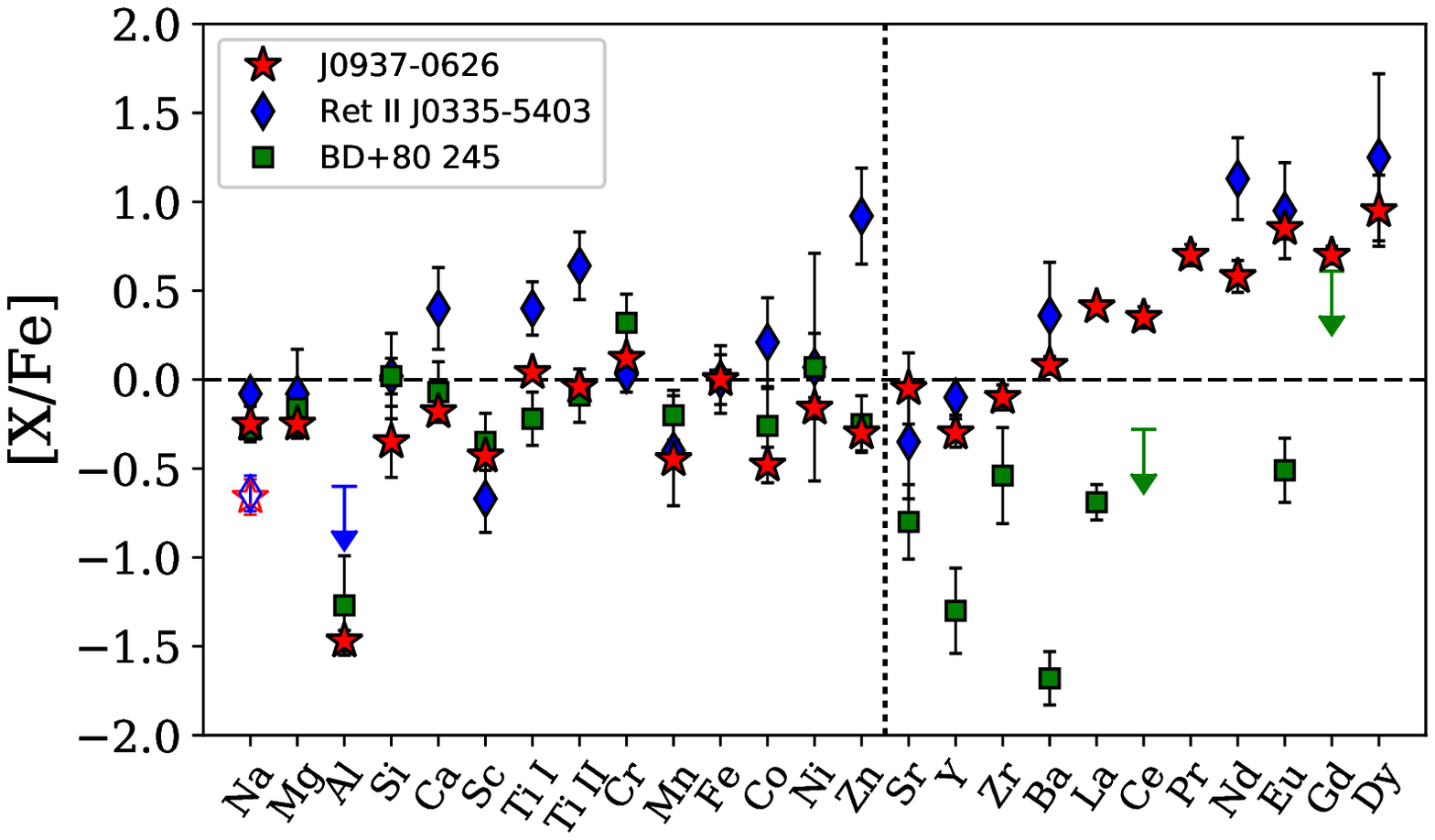}
\caption{[X/Fe] ratios for each element in \targetsp (red stars),
  \RetIIStarsp (blue diamonds; \citealt{Ji2016}), and BD$+80^{\circ}$~245
(green squares; \citealt{Ivans2003} for Na, Si, La, and Eu,
  \citealt{Roederer2014b} for all other elements).  NLTE-corrected Na
  abundances for \targetsp and \RetIIStarsp are shown with open
  symbols; NLTE-corrected values are not shown for Mn and Co.  The
  neutron-capture elements are on the right, separated by a dotted
  line.
\label{fig:DelXFe}}
\end{center}
\end{figure}

A more detailed element-by-element comparison is shown in Figure
\ref{fig:DelXFe}, which plots [X/Fe] ratios for each element.
Following \citet{Yong2013}, only stars with similar metallicities are
shown---here, \targetsp ($[\rm{Fe/H}]=-1.86$) is shown along with
\RetIIStarsp ($[\rm{Fe/H}]=-2.19$; \citealt{Ji2016}) and
BD$+80^{\circ}$~245 ($[\rm{Fe/H}]=-2.04$;
\citealt{Ivans2003,Roederer2014b}).  For the elements through Zn,
\target's abundance ratios are similar to BD$+80^{\circ}$~245.
\RetIIStarsp shows a similar abundance pattern, with the exception of
Ca, Ti, Co, and Zn, which all have higher [X/Fe] ratios than \target.

To investigate the source of \target's unusual abundance ratios, the
procedure from \citet{McWilliam2018}, who performed a re-analysis
of UMi COS~171, is followed.  First, note that the removal of $\sim 0.6$
dex of Fe from \targetsp would shift its $[\alpha/\rm{Fe}]$ ratios to
normal values---this shift would lead to a lower metallicity of
$[\rm{Fe/H}]=-2.50$.  The specific yields from the event(s) that
created \target's unusual abundance pattern can then be investigated
through a comparison with a star at $[\rm{Fe/H}]=-2.50$ that formed in
the same environment. \citet{McWilliam2018} used a more metal-poor
star in UMi; however, such a star cannot be confidently identified for
\target. Instead, the average abundance pattern of MW stars with
$-2.6~\le~[\rm{Fe/H}]~\le~-2.4$ is utilized.  To make this comparison
as homogeneous as possible, the average abundances of the MW stars in
this metallicity range from \citet{Sakari2018b} are used. These stars
have been analyzed with the same techniques as \target, using the same
$<$3D$>$, NLTE corrections to \ion{Fe}{1} lines, the same model
atmospheres, and the same line lists.  (Though note that none of the
stars in \citealt{Sakari2018b} are as hot as \target.)  These average
values are also shown in Figures \ref{fig:alphaFe} and
\ref{fig:FePeakFe}. Note that the NLTE corrections to the Mn and Co
abundances will differ slightly between $[\rm{Fe/H}]=-1.86$ and
$[\rm{Fe/H}]=-2.5$---however, the MPIA NLTE correction database
indicates that the corrections are $\la 0.1$ dex higher at
$[\rm{Fe/H}]=-2.5$ for both Mn and Co
\citep{BergemannGehren2008,Bergemann2010}.

\begin{figure}[h!]
\begin{center}
\centering
\hspace*{-0.3in}
\includegraphics[scale=0.8,clip,trim=0.0in 0 0 0]{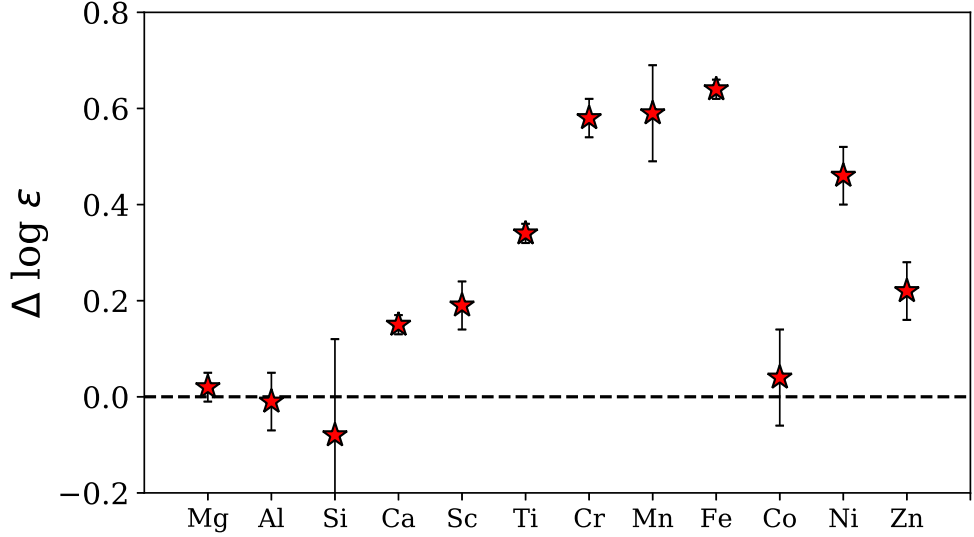}
\caption{Differences in $\log \epsilon$ abundance between \targetsp
  and a typical MW star at $[\rm{Fe/H}] = -2.5$ (the Fe abundance that
  \targetsp would need to have normal $[\alpha/\rm{Fe}]$ ratios).
  \ion{Ti}{2} and \ion{Cr}{2} are chosen to represent Ti and Cr, to
  minimize NLTE effects.  Note that LTE abundances are shown, though
  both Mn and Co require NLTE corrections; however, these corrections
  differ by $\la -0.1$ between stars at $[\rm{Fe/H}] = -1.86$ and
  $[\rm{Fe/H}] = -2.5$, according to the MPIA NLTE database.
\label{fig:DelLogEps}}
\end{center}
\end{figure}

Figure \ref{fig:DelLogEps} then shows the differences in the
$\log\epsilon$ abundance between \targetsp and the average MW values
at $[\rm{Fe/H}]=-2.5$.  This comparison indicates that one or more
nucleosynthetic events have significantly enriched \targetsp in Cr,
Mn, Fe, and Ni, with minor enhancement in Ca, Sc, Ti, and Zn.

\subsection{Potential Explanations for Light, $\alpha$, and Fe-peak Abundances}\label{subsec:Scenarios}
\targetsp therefore either shows an enhancement in some Fe-peak
elements relative to the light and $\alpha$-elements, or a relative
deficiency in the light and $\alpha$-elements.  Three possible
scenarios to explain this abundance pattern are considered. First, the
abundances could reflect an evolutionary effect, such as radiative
levitation.  Secondly, the entire abundance pattern could be
representative of enrichment from a single object. Finally, the
abundance profile could be due to multiple progenitors, as a
consequence of extended star formation.  These possibilities are
addressed below.

Radiative levitation in hot horizontal branch stars has been shown to
enhance Fe-peak abundances by large amounts (up to 3 dex;
\citealt{Behr2003}). Reproducing the observed [X/Fe] ratios would
require levitation only for the Fe-peak (and neutron-capture)
elements.  Furthermore, significant abundance differences have only
been observed in the hottest horizontal branch stars, with
temperatures above $\sim~11,000$~K \citep{Lovisi2012,Tailo2017}; at
5875 K, \targetsp is not expected to experience significant radiative
levitation. Such effects are also not seen in other field horizontal
branch stars (e.g., HD~222925; \citealt{Roederer2018b}). This scenario
therefore seems unlikely to explain the abundance pattern in \target.

Enrichment by a single source has also been invoked as an explanation
for $\alpha$-poor very metal-poor stars.  Standard core-collapse
supernovae are unlikely to produce sufficiently low [$\alpha$/Fe]
ratios to match those in \target; however, more exotic supernovae can
create unusual abundance signatures. \citet{Aoki2014} found that a
pair-instability (PISN) supernova could explain the abundance
signature of SDSS~J0018$-$0939, a star at
$[\rm{Fe/H}]~\sim~-2.5$. They based this conclusion on the star's low
[$\alpha$/Fe] (with the exception of Si), [C/Fe], and [Co/Fe] ratios, 
as well as its strong odd-even effect (contrasting abundances in odd
vs. even elements), though they do note that the predicted odd-even
effect is stronger than observed.   \citet{Nagasawa2018} also explore
the possibility that a PISN supernova enriched their three stars in
Hor~I (with metallicities ranging from $[\rm{Fe/H}]=-2.8$ to $-2.5$),
finding that none of the PISN models can perfectly reproduce the
abundance pattern.  They note that the models do not match the
observed Fe-peak abundances (particularly their solar [Co/Fe] ratios)
and predict a strong odd-even effect that is not observed.  \targetsp
also has low [$\alpha$/Fe] and [Co/Fe]---however, the
\citet{HegerWoosley2002,HegerWoosley2010} PISN models do not match all
the abundance ratios.  In particular, a PISN supernova cannot produce
enough Sc or Zn, and produces a stronger odd-even effect than
observed.  It therefore seems unlikely that \targetsp was enriched by
a single PISN.

\citet{Nishimura2017} and \citet{TsujimotoNishimura2018} have noted
that neutrino-heating in magneto-rotational supernovae may be a viable
site for Zn production (along with Fe, Co, and Ni) in metal-poor
environments.  Using samples of stars in the Milky Way,
\citet{TsujimotoNishimura2018} argue that a high [Zn/Mg] ratio at very
low metallicity indicates a high frequency of magneto-rotational
supernovae; at higher metallicities, a high [Zn/Mg] may also reflect
the onset of Type Ia supernovae (see their Figure 3).  Given its high
$[\rm{Zn/Mg}]=-0.05$ (compared to an average $[\rm{Zn/Mg}]~=~-0.40$ at
$[\rm{Fe/H}]=-1.9$), it is tempting to speculate that \targetsp may
have been enriched by a magneto-rotational supernova; however, by
$[\rm{Fe/H}] = -2.5$, the contributions from magneto-rotational
supernovae are already expected to be decreasing.  Instead,
\citet{TsujimotoNishimura2018} argue that the Zn enhancement in
\RetIIStarsp and a star in the Draco dwarf galaxy is due to Type Ia
supernovae.  Detailed yields from magneto-rotational supernovae are
necessary to fully address this possibility.

\citet{Kobayashi2014} have also argued that low [$\alpha$/Fe] in
extremely metal-poor stars can be the nucleosynthetic result of
$\sim10-20$ M$_{\sun}$ core-collapse supernovae or hypernovae.  They
suggest that hypernovae would produce a high $[\rm{Zn/Fe}]>0.3$, which
is not observed in \target.  The abundance pattern for $\sim10-20$
M$_{\sun}$ supernovae also does not quite match the pattern in
\target, particularly the high $\alpha$ and Zn and the pattern of
Fe-peak elements.  Core-collapse supernovae on their own are
therefore not a likely source of the abundance patterns in \target.
It is also worth noting that \targetsp is more metal-rich than the
other targets whose abundances were explained by a single progenitor.

A more likely explanation for the abundance patterns in Figures
\ref{fig:DelXFe} and \ref{fig:DelLogEps} is that \target's host
environment experienced extended star formation and chemical
enrichment, with core-collapse supernovae building the metallicity up
to $[\rm{Fe/H}] = -2.5$ before a second event produced a significant
amount of Cr, Mn, Fe, and Ni and a smaller amount of Ca, Sc, Ti, and
Zn.  The most likely option is enrichment from a Type Ia supernova,
which are known to produce Fe-peak elements (e.g.,
\citealt{Iwamoto1999,Badenes2003,Badenes2008}).  The precise yields
depend on parameters such as white dwarf mass, metallicity, and the
physics of the explosion.

\begin{figure}[h!]
\begin{center}
\centering
\hspace*{-0.3in}
\includegraphics[scale=0.8,clip,trim=0.0in 0 0 0]{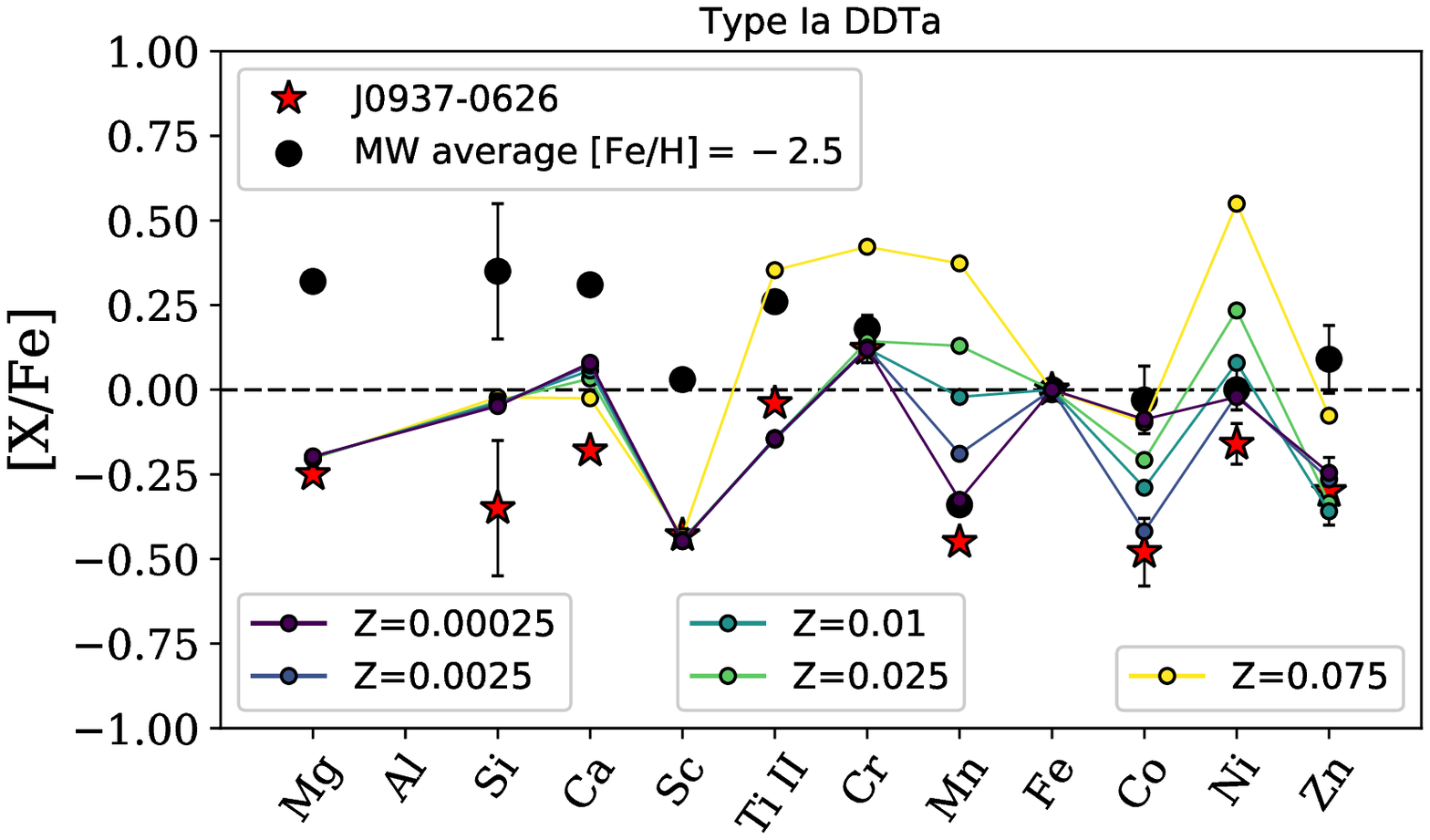}
\caption{Comparisons of ``DDTa'' Ia yields from \citet[as reported by
    \citealt{McWilliam2018}]{Badenes2003,Badenes2008} with the
  \targetsp abundances (red stars).  The Ia yields are added to the
  average MW abundances at $[\rm{Fe/H}] \sim -2.5$ (black circles) so
  that the [Fe/H] is increased to \target's value.  The five lines
  show the Ia yields for five different metallicities. Note that the
  yield patterns change when the physics of the explosion are altered;
  the ``DDTa'' model represents the best fit to \target's abundances.
\label{fig:IaYields}}
\end{center}
\end{figure}

\begin{figure}[h!]
\begin{center}
\centering
\subfigure{\includegraphics[scale=0.7,trim=0.02in 0.1in 0.25in 0.0in,clip]{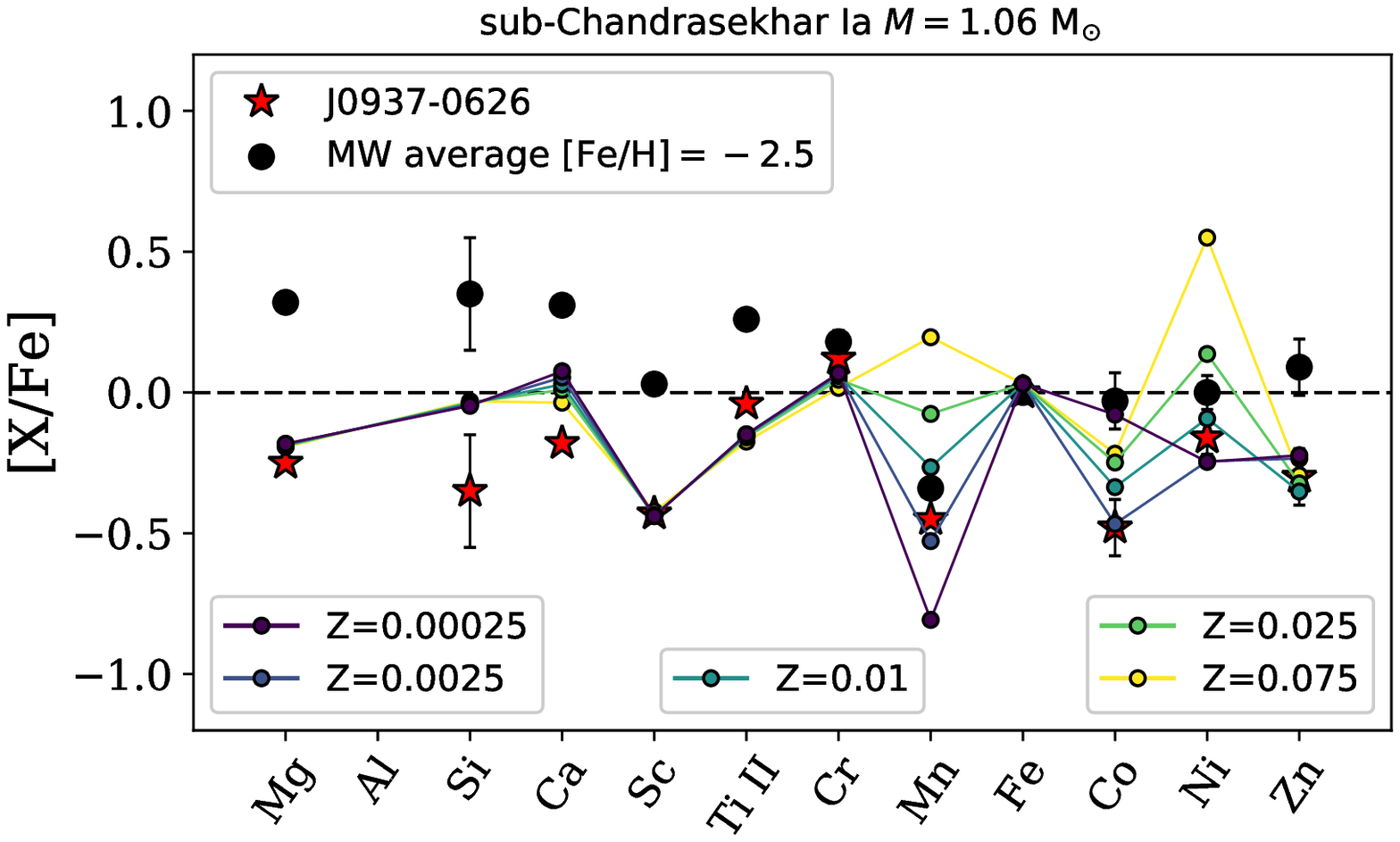}\label{subfig:subCh106}}
\subfigure{\includegraphics[scale=0.7,trim=0.02in 0.1in 0.25in 0.0in,clip]{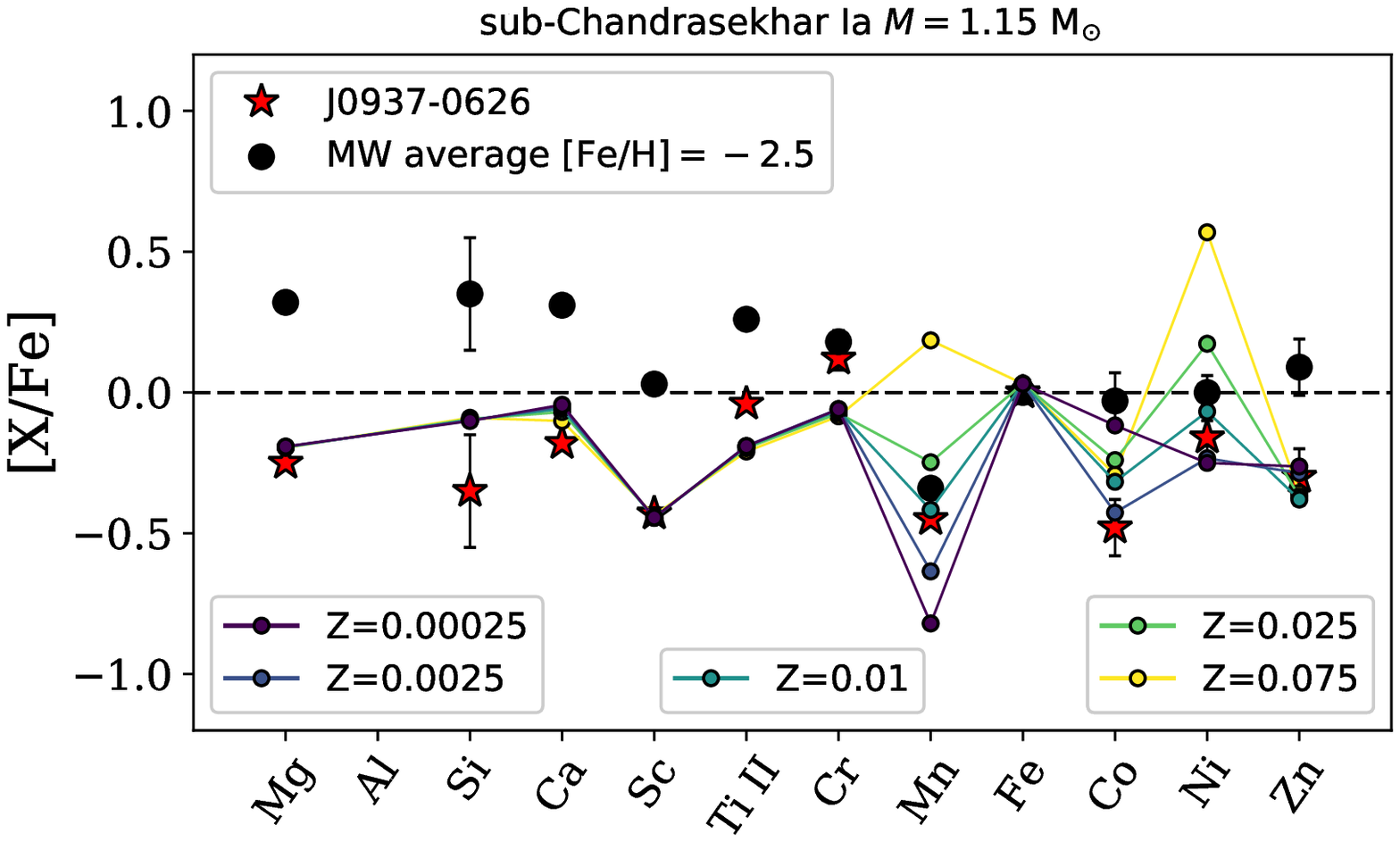}\label{subfig:subCh115}}
\caption{Comparisons of sub-Chandrasekhar mass Ia yields from
  E. Bravo, as reported by \citet{McWilliam2018} with \targetsp
  abundances.  Points are as in Figure \ref{fig:IaYields}.
\label{fig:subChIaYields}}
\end{center}
\end{figure}

Figures \ref{fig:IaYields} and \ref{fig:subChIaYields} show
comparisons between \target's [X/Fe] abundance ratios and Ia yields,
added to the background MW average at $[\rm{Fe/H}]\sim -2.5$.  The
Chandrasekhar-mass ``DDTa'' model from \citet{Badenes2003,Badenes2008}
is shown in Figure \ref{fig:IaYields}, using the yields from
\citet{McWilliam2018}, for five different metallicities.  (Note that
the other models overpredict [Mn/Fe].)  Though the agreement for
$Z=0.0025$ is generally decent, none of these metallicities can
perfectly reproduce the pattern in the Fe-peak elements.  At all
metallicities, this Ia model also overproduces Si and Ca and
(except for the highest metallicity model) underproduces Ti.  Figure
\ref{fig:subChIaYields} then shows sub-Chandrasekhar mass models from
E. Bravo, with the yields from \citet{McWilliam2018}, for two white
dwarf masses: 1.06~M$_{\sun}$ and 1.15~M$_{\sun}$.  For UMi COS~171,
\citet{McWilliam2018} found that a sub-Chandrasekhar mass model
provided a better fit to the abundances, particularly the low [Mn/Fe]
and [Ni/Fe].  Indeed, there is decent agreement with Mn and Co in
\targetsp for both sub-Chandrasekhar masses.  As with the Ia case,
both sub-Chandrasekhar models overpredict the amount of Si and Ca.

Though no model perfectly matches the pattern in \target, its
abundances are generally consistent with enrichment from a Type Ia
supernova, possibly one with a sub-Chandrasekhar mass.  However, the
precise Ia yields depend on the physical conditions of the models
(e.g., explosion energy).  Similarly, if the background composition of
\target's birth environment was different from the MW composition
(e.g., if the Si and Ca abundances were lower at
$[\rm{Fe/H}]~=~-2.5$), then these predicted yields would also change.
The general enhancement in Fe-peak elements supposts enrichment from a
Type Ia supernova.

It is worth noting that any of the proposed scenarios for enrichment
in Fe-peak elements are unlikely to have created a significant number
of neutron-capture elements.  Neither Type Ia or PISN supernovae will
create $r$-process elements (e.g., \citealt{HegerWoosley2002}), while
standard core-collapse supernovae have been ruled out as a significant
source of $r$-process elements (e.g.,
\citealt{ArconesThielemann2013}).  Though magneto-rotational
supernovae have been identified as possible sources of both Fe-peak
and $r$-process elements, \citet{Nishimura2017} showed that the
supernovae that produce significant amounts of Fe, Ni, and Zn do not
produce much Eu, and vice versa.  The enhancement of neutron-capture
elements therefore likely requires enrichment by a separate event.

\clearpage

\subsection{r-Process Enhancement and Patterns}\label{subsec:Patterns}
\begin{figure*}[h!]
\begin{center}
\centering
\includegraphics[scale=0.75,trim=0.05in 0.4in 0.55in 0.8in,clip]{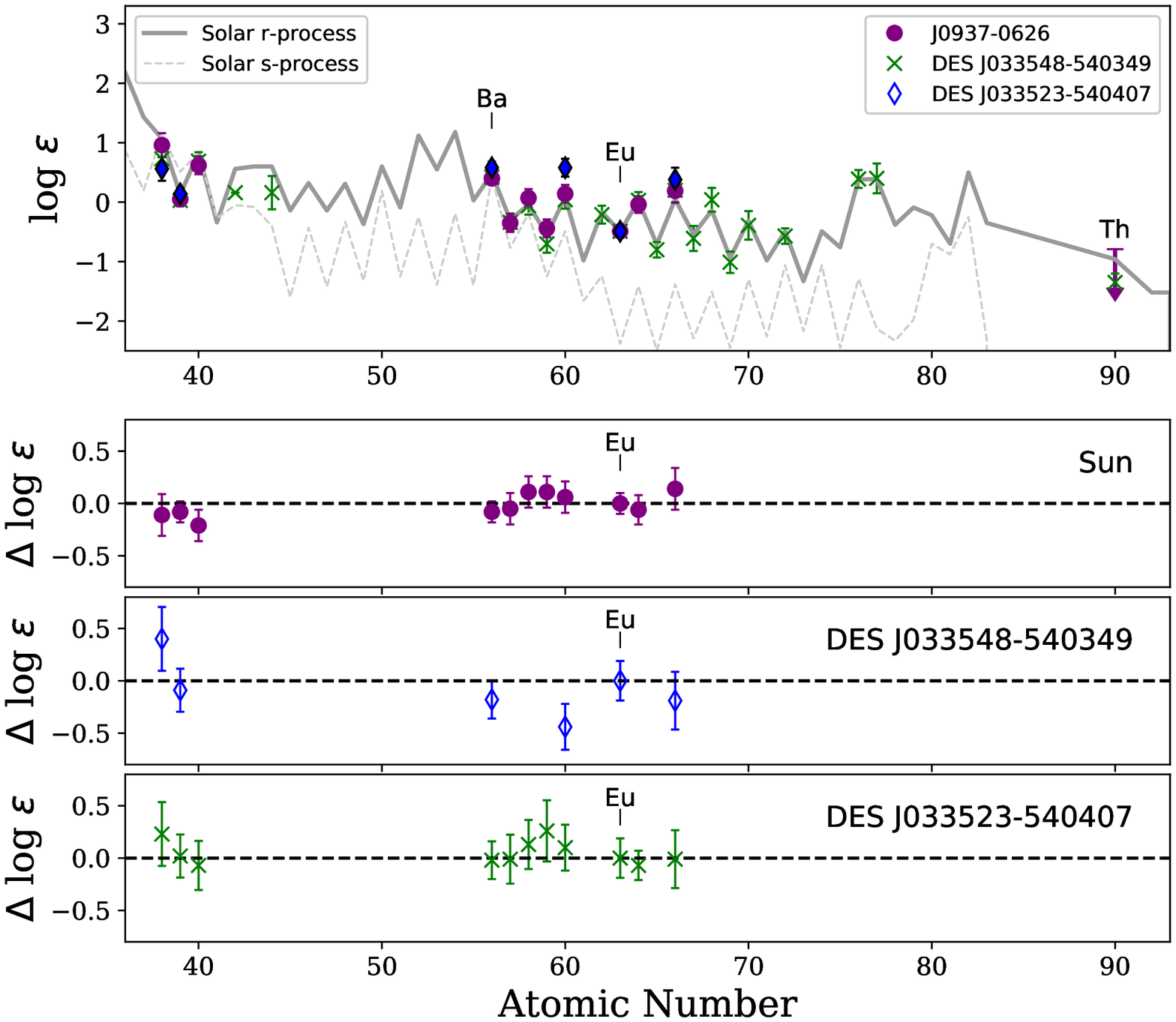}
\caption{{\it Top: } Abundances of neutron-capture elements in
  \targetsp along with the total errors (Table \ref{table:Abunds});
  also shown are the $r$- and $s$-process patterns in the Sun (grey
  line, from \citealt{Arlandini1999}) and two stars in Ret~II
  \citep{Ji2016}.  The solar $r$-process pattern and the Ret II
  abundances are shifted to the Eu abundance in \target; the solar
  $s$-process pattern is shifted to match the Ba abundance.  {\it Bottom
    three panels: } The residuals between \targetsp and the Sun, DES
  J033548$-$540349, and DES J033523$-$540407.\label{fig:rProcPatterns}}
\end{center}
\end{figure*}

Unlike many of the light, $\alpha$, and Fe-peak elements, \targetsp
shows solar or supersolar [X/Fe] ratios for the neutron-capture
elements. At $[\rm{Eu/Fe}] = 0.85\pm0.06$, \targetsp is an
$r$-I star; its low $[\rm{Ba/Eu}]=-0.77\pm0.07$ implies that its Eu
enhancement is due to the $r$-process.  Note that red horizontal
branch stars have been discovered to be $r$-process enhanced
\citep{Roederer2014n}, including HD~222925 \citep{Roederer2018b}, so
this is not a unique feature of \target.  Sr, Y, and Zr are also
roughly solar, while La, Ce, Pr, Nd, Gd, and Dy are enhanced. Figure
\ref{fig:rProcPatterns} shows that \target's neutron-capture abundance
pattern is generally consistent with the $r$-process residual in the
Sun and two stars in Reticulum~II (though, like Ret~II, Sr, Y, and Zr
in \targetsp are slightly lower than the solar residual, as discussed
in \citealt{Ji2016}).  The pattern is inconsistent with the solar
$s$-process.  The low [Sr/Ba] also indicates that significant
contributions from the weak $s$-process in rapidly rotating massive
stars (e.g.,
\citealt{Chiappini2011,Frischknecht2012,Cescutti2013,Frischknecht2016})
are unlikely.  The upper limit in Th also implies
$\log\epsilon(\rm{Th/Eu})~<~-0.3$.

\begin{figure}[h!]
\begin{center}
\centering
\hspace*{-0.45in}
\subfigure{\includegraphics[scale=0.58,trim=0.02in 0.1in 0.55in 0.4in,clip]{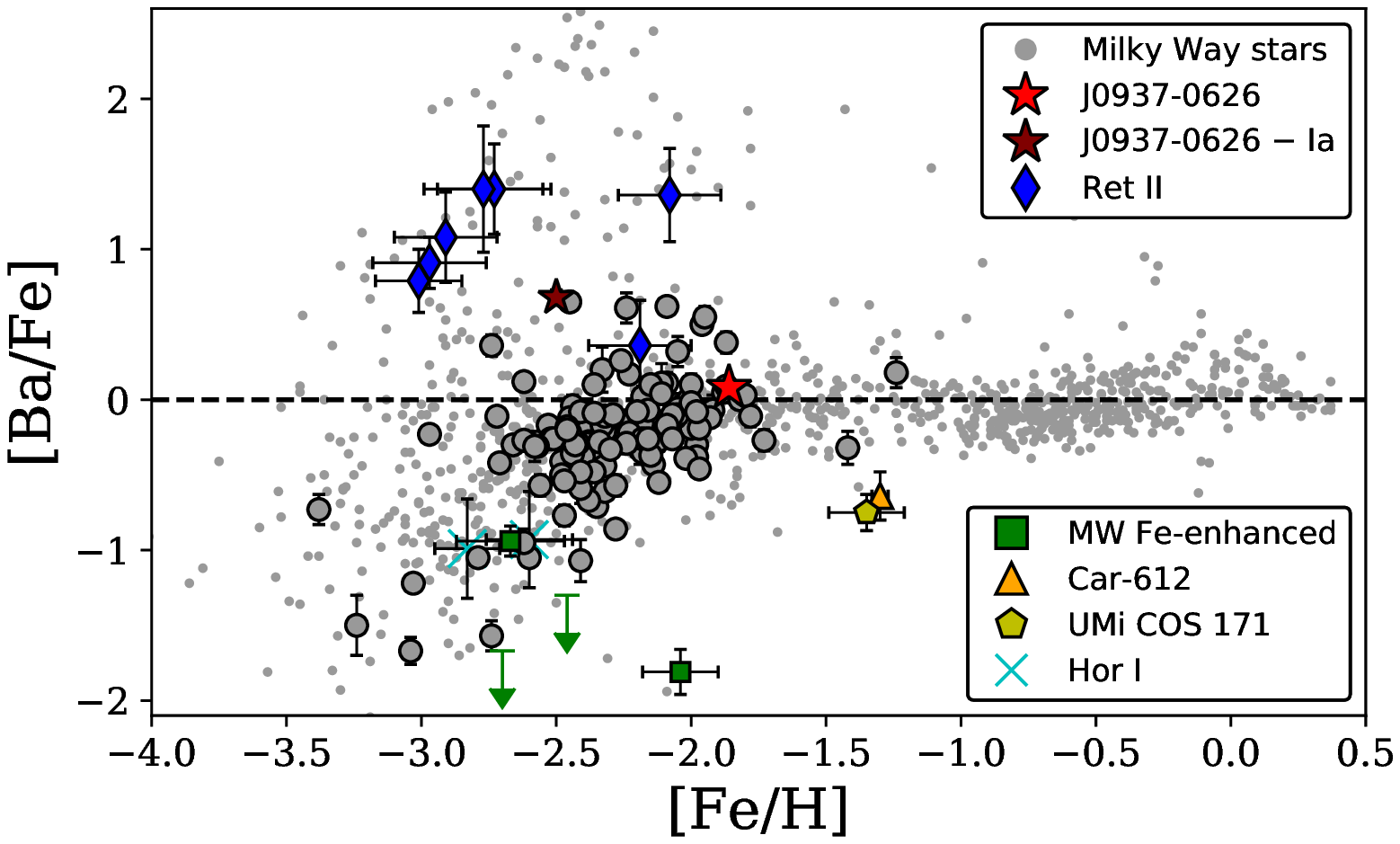}\label{subfig:Ba}}
\subfigure{\includegraphics[scale=0.58,trim=0.02in 0.1in 0.55in 0.4in,clip]{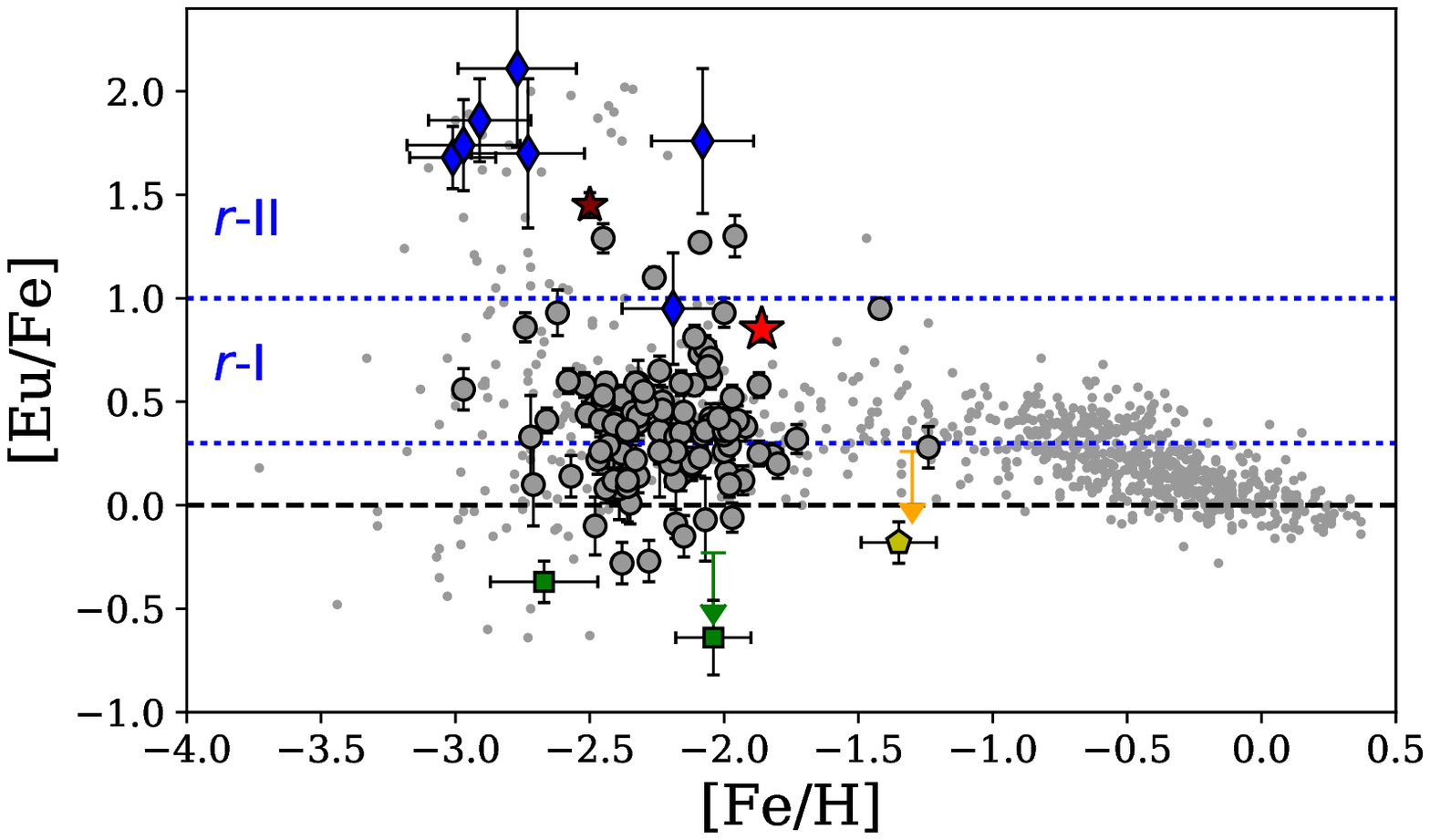}\label{subfig:Eu}}
\caption{[Ba/Fe] and [Eu/Fe] as a function of [Fe/H].  The points are
  as in Figure \ref{fig:alphaFe}, except that the MW average is not
  shown.  Instead, the maroon star shows how \target's abundance
  ratios would change with the removal of 0.6 dex of Fe.  The right
  panel also shows the $r$-I and $r$-II [Eu/Fe] definitions.
\label{fig:NCFe}}
\end{center}
\end{figure}

Figure \ref{fig:NCFe} shows \target's slightly enhanced [Ba/Fe] and
strongly enhanced [Eu/Fe], relative to the ``normal'' and
``Fe-enhanced'' MW field stars.  The $r$-process enhancement in
\targetsp makes it unlike most of the other ``Fe-enhanced'' metal-poor
stars, whose low [X/Fe] ratios persist through the neutron-capture
elements (including BD$+80^{\circ}$~245; see Figure
\ref{fig:DelXFe}).\footnote{Note that though there are
  $r$-process-enhanced stars in UMi, COS~171 is not
  $r$-process-enhanced.} Instead, the $r$-process enhancement in
\targetsp more closely resembles \RetIIStar, the Ret~II star, which is
also an $r$-I star.

Section \ref{subsec:MWComp} demonstrated that many of the [X/Fe]
ratios could be brought into agreement with Milky Way stars by
removing 0.6 dex of Fe.  The removal of 0.6 dex of Fe would
increase the [X/Fe] ratios of the $r$-process elements, as shown by
the maroon star in Figure \ref{fig:NCFe}.  If the $r$-process event
occurred prior to the Fe-peak event, then \targetsp would have been an
$r$-II star if the Fe-peak event had not occurred. Furthermore, if
\targetsp originated in a dwarf galaxy (see Section
\ref{subsec:Kinematics}), this dwarf galaxy would likely have
contained a population of highly $r$-process-enhanced stars, similar
to Ret~II. Indeed, though none have been linked to \target, many of
the $r$-II stars in the MW have been kinematically identified as
probable captures from dwarf galaxies
\citep{Roederer2018,Sakari2018a,Sakari2018b}, hinting that many $r$-II
stars have originated in $r$-process-enhanced dwarf galaxies like
Ret~II.

\clearpage

\subsection{Kinematics}\label{subsec:Kinematics}
\begin{figure}[h!]
\begin{center}
\centering
\includegraphics[scale=0.65,trim=0.2in 0.0in 0.0in 0.5in,clip]{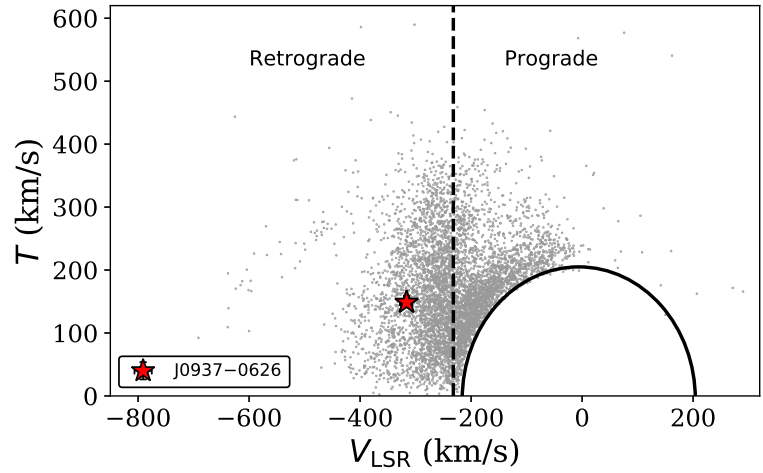}
\caption{A Toomre diagram, where $T = \sqrt{U^2+W^2}$, utilizing
  parallaxes and proper motions from {\it Gaia} DR2.  \targetsp is
  shown as a red star. The grey points are MW halo stars within
  1~kpc, from \citet{Koppelman2018}; the large black circle shows
  their definition for halo membership, where disk stars lie within
  the circle.
  \label{fig:Toomre}}
\end{center}
\end{figure}

Figure \ref{fig:Toomre} shows a Toomre diagram of MW field stars from
{\it Gaia} DR2 (using the halo stars within 1 kpc from
\citealt{Koppelman2018}), along with distinctions between prograde and
retrograde orbits. Its velocities (derived with the \texttt{gal\_uvw}
code\footnote{\url{https://github.com/segasai/astrolibpy/blob/master/astrolib/gal_uvw.py}})
show that \targetsp resides in the MW halo, with a retrograde
orbit.  Previous work has found that a significant number of MW $r$-I
and $r$-II stars have retrograde orbits
\citep{Roederer2018,Sakari2018a,Sakari2018b} and may have been
accreted from satellite galaxies.  \target's kinematics also
suggest that it may also have been accreted from a satellite galaxy.

\citet{Koppelman2018} and \citet{Roederer2018} have also identified
specific groups of stars with similar kinematics which may have
originated in the same galaxy.  Recently, \citet{Helmi2018} argued
that the majority of the retrograde stars from Koppelman et al.'s
analysis are due to a single merger event from a galaxy with a mass
slightly higher than the Small Magellanic Cloud, which they named
Gaia-Enceladus.  They also found that the Gaia-Enceladus stars have
slightly lower [$\alpha$/Fe] ratios than MW stars (also see
\citealt{NissenSchuster2010} and \citealt{Hayes2018}).  \targetsp lies
approximately in the correct kinematic space for Gaia-Enceladus stars;
however, its [Fe/H] and [$\alpha$/Fe] ratios are lower than the
majority of the Gaia-Enceladus stars.  It is still possible that
\targetsp was brought in by the same merger event if it experienced
inhomogeneous mixing within the larger galaxy (similar to the scenario
proposed by \citealt{Venn2012} for Carina).  Full orbital calculations
will also be essential for identifying \target's birth site and
locating other stars from the same environment.

\citet{Roederer2018} examined the kinematics of 35 $r$-II stars with
high-quality {\it Gaia} data, and identified several groups with
similar orbits and metallicities.  This technique could be used to
identify other stars from the same birth environment as \target: its
chemistry and kinematics should be similar to other $r$-I stars from
the same birth environment; similarly, if that environment was
enriched in $r$-process elements before the event that created the
Fe-peak enrichment, \targetsp should have similar kinematics as more
metal-poor $r$-II stars.  RPA discoveries of more $r$-I and $r$-II
stars, combined with future {\it Gaia} data, will identify other stars
that could have originated in the same environment as \target.

\clearpage


\section{Conclusions}\label{sec:Conclusions}
\longtarget $\;$ is a moderately $r$-process-enhanced
($[\rm{Eu/Fe}]~=~+~0.85\pm0.1$), metal-poor ($[\rm{Fe/H}]~=~-1.86$)
horizontal branch star on a retrograde orbit in the MW halo that was
identified by the RPA.  Most of its [X/Fe] abundance ratios are
distinct from those of typical MW field stars, particularly its
subsolar $[\alpha/\rm{Fe}]$ (e.g., $[\rm{Mg/Fe}]~=~-~0.25\pm0.04$,
$[\rm{Ca/Fe}]~=~-0.18\pm0.03$), light element
($[\rm{Na/Fe}]~=~-0.25\pm0.13$) and some Fe-peak ratios (e.g.,
$[\rm{Ni/Fe}]~=~-0.18\pm0.06$).  \targetsp seems to have the abundance
pattern typical of a ``normal'' MW star at $[\rm{Fe/H}]\sim-2.5$ 
that was diluted by ejecta from an event that created $\sim 0.6$ dex
of Fe-peak elements. Although none of the models perfectly fit the
abundance patterns in \target, the best candidate for this Fe-peak
enrichment is a Type Ia supernova.

\target's $r$-process enrichment is unlikely to have been caused by
a Type Ia supernova. Instead, its birth environment may have been
enhanced in $r$-process elements prior to the enrichment from the Type
Ia supernova; \targetsp therefore could have been an $r$-II star were
it not for the occurrence of the Type Ia supernova. In this sense,
\targetsp may be similar to the metal-rich $r$-I star in Reticulum~II.
Ultimately, \target's chemical abundances and kinematics indicate that
it was likely accreted from a satellite dwarf galaxy.  \target's host
galaxy may have been responsible for depositing other stars into the
MW halo, possibly even more metal-poor $r$-II stars.  Additional
discoveries of $r$-I and $r$-II stars by the RPA, combined with proper
motions and parallaxes from {\it Gaia}, will enable specific subgroups
to be identified in the future.

\acknowledgements
The authors thank the current and previous observing specialists on
the 3.5-m telescope at Apache Point Observatory for their continued
help and support.  The authors thank the anonymous referee for helpful
comments which improved this manuscript.  The authors also thank Anish
Amarsi and Karin Lind for providing the NLTE grids and assisting with
their usage, and Helmer Koppelman for providing the 1 kpc {\it Gaia}
sample.

C.M.S. and G.W. acknowledge funding from the Kenilworth Fund
of the New York Community Trust.
V.M.P., T.C.B., R.E., A.F., and I.U.R. acknowledge partial support
from grant PHY 14-30152 (Physics Frontier Center/JINA/CEE), awarded by
the US National Science Foundation.   I.U.R. acknowledges additional
support from NSF grants AST~16-13536 and AST~18-15403.
BKG acknowledges the support of STFC, through the University of
Hull's Consolidated Grant ST/R000840/1.
EKG gratefully acknowledges support via Sonderforschungsbereich SFB
881 ``The Milky Way System'' (subproject A5) of the German Research
Foundation (DFG).  Funding for RAVE has been provided by: the
Leibniz-Institut f\"{u}r Astrophysik Potsdam (AIP); the Australian
Astronomical Observatory; the Australian National University; the
Australian Research Council; the French National Research Agency; the
German Research Foundation (SPP 1177 and SFB 881); the European
Research Council (ERC-StG 240271 Galactica); the Istituto Nazionale di
Astrofisica at Padova; The Johns Hopkins University; the National
Science Foundation of the USA (AST-0908326); the W. M. Keck
foundation; the Macquarie University; the Netherlands Research School
for Astronomy; the Natural Sciences and Engineering Research Council
of Canada; the Slovenian Research Agency; the Swiss National Science
Foundation; the Science \& Technology Facilities Council of the UK;
Opticon; Strasbourg Observatory; and the Universities of Groningen,
Heidelberg and Sydney. The RAVE web site is at
\url{https://www.rave-survey.org}.

This research is based on observations obtained with the Apache Point
Observatory 3.5-meter telescope, which is owned and operated by the
Astrophysical Research Consortium.  This research has made use of the
SIMBAD database, operated at CDS, Strasbourg, France. This work has
also made use of data from the European Space Agency (ESA) mission
{\it Gaia} (\url{http://www.cosmos.esa.int/gaia}), processed by the
{\it Gaia} Data Processing and Analysis Consortium (DPAC,
\url{http://www.cosmos.esa.int/web/gaia/dpac/consortium}).  Funding
for the DPAC has been provided by national institutions, in particular
the institutions participating in the {\it Gaia} Multilateral
Agreement.

\software{IRAF (Tody 1986, Tody 1993), DAOSPEC
  \citep{DAOSPECref}, MOOG (v2017; \citealt{Sneden,Sobeck2011}),
 linemake (\url{https://github.com/vmplacco/linemake}), gal\_uvw (\url{https://github.com/segasai/astrolibpy/blob/master/astrolib/gal_uvw.py})}

\footnotesize{

}

\end{document}